\documentclass[useAMS, usenatbib]{mn2e}
\usepackage{aas_macros}
\usepackage{amsmath}
\usepackage{amssymb} 
\usepackage{graphicx}
\usepackage{subcaption} 
\usepackage{float}
\usepackage{booktabs} 
\usepackage{multirow} 
\usepackage{amsfonts}
\usepackage{lmodern}
\usepackage{bm}
\usepackage{upgreek}

\newcommand{\bc}{\begin{center}}
\newcommand{\ec}{\end{center}}

\renewcommand{\vec}[1]{\mathbf{#1}}
\let\oldhat\hat
\renewcommand{\hat}[1]{\oldhat{\mathbf{#1}}}
\let\oldbullet\bullet \renewcommand{\bullet}[1][0pt]{%
\mathrel{\raisebox{#1}{$\oldbullet$}}%
}

\title{Resolving flows around black holes: the impact of gas angular momentum} \author[Curtis \& Sijacki]{Michael Curtis$^1$\footnotemark[1], Debora
  Sijacki$^{1}$ \\
  $^1$ Institute of Astronomy and Kavli Institute for Cosmology,
  University of Cambridge, Madingley Road, Cambridge CB$3$ $0$HA, UK}

\begin{document}

\maketitle

\begin{abstract}

Cosmological simulations almost invariably estimate the accretion of gas on to
supermassive black holes using a Bondi-Hoyle-like prescription. Doing so
ignores the effects of the angular momentum of the gas, which may prevent or
significantly delay accreting material falling directly on to the black
hole. We outline a black hole accretion rate prescription using a modified
Bondi-Hoyle formulation that takes into account the angular momentum of the
surrounding gas. Meaningful implementation of this modified Bondi-Hoyle
formulation is only possible when the inner vorticity distribution is well
resolved, which we achieve through the use of a super-Lagrangian refinement
technique around black holes within our simulations. We then investigate
the effects on black hole growth by performing simulations of isolated as well
as merging disc galaxies using the moving-mesh code {\small AREPO}. We find that
the gas angular momentum barrier can play an important role in limiting the
growth of black holes, leading also to a several Gyr delay between the
  starburst and the quasar phase in major merger remnants. We stress, however,
that the magnitude of this effect is highly sensitive to the thermodynamical state of
the accreting gas and to the nature of the black hole feedback present.

\end{abstract}


\section{Introduction}
\renewcommand{\thefootnote}{\fnsymbol{footnote}}
\footnotetext[1]{E-mail: mc636@ast.cam.ac.uk}

Gas falling on to supermassive black holes that reside in the centres of
majority galaxies may have a significant amount of angular momentum with
respect to the central black hole. The accreting gas will settle into orbits
that are aligned with a plane normal to the mean angular momentum of the
infalling gas. Once the fluid has settled on to circular orbits accretion on to
the central object will be inhibited by the centrifugal force and, as such, a
disc like structure will form, with the subsequent evolution of the system
governed by, amongst others, the viscous processes that are present in the disc \citep{Pringle:81, Frank:02}. The viscosity of the fluid causes inner packets of fluid to move inwards and, in doing so, they transfer angular momentum to the outer parts of the disc. This process causes most of the mass to spiral into the central object, whilst transporting angular momentum outwards.

One key problem with the predictive power of accretion disc theory is understanding the nature and relative importance of the different sources of viscosity. If we assume that the source of the viscosity is molecular then we can estimate the ratio of the inertial to viscous forces, the Reynolds number, as
\begin{equation}
\mathit{Re} = \frac{u_\phi R}{\nu}\,,
\end{equation}
where $u_\phi$ is the rotational velocity of the fluid, $\nu \sim
c_\mathrm{s}\lambda$ is the molecular viscosity of the fluid, $c_\mathrm{s}$
is the sound speed of the gas and $\lambda$ is the mean free path
length. Putting in typical values for accretion discs yields a Reynolds number
$\sim 10^{15}$, which implies an accretion timescale larger than a Hubble
time. Clearly, to explain the accretion rates on to observed active galactic
nuclei (AGN), we need to invoke different viscosity mechanisms. These are
likely to include magnetic effects, especially at small spatial scales, such
as the magnetic rotational instability, which occurs in sufficiently
ionized flows for which $u_\phi$ decreases as a function of radius
\citep{Balbus:91}. In such cases, the effect is to negatively torque inner
fluid elements, causing them to fall on to closer orbits, and vice versa. This
motion is unstable and leads to a magnetic viscosity. Another source of
viscosity may be caused by the turbulent motion of the disc, where the random
flow and eddies of the turbulence lead to a viscosity analogous to that
driven by random flow on the molecular level. The nature of such turbulence
and at what point in the flow it sets in remains, however, a key uncertainty.

As there are still fundamental gaps in our understanding of gas angular
momentum transport, we are limited in our ability to successfully model black hole
accretion discs. Hence, much modern theory follows \citet{Shakura:73} in
parameterizing the viscosity as $\nu=\alpha c_\mathrm{s}H$, where $H$ is the
disc scaleheight and $0<\alpha \lesssim 1.0$ is a dimensionless free
parameter that essentially represents our ignorance about the exact nature of
the viscosity. Within this framework one can solve the steady disc equations
for the geometrically thin disc regime. While widely used this approximation
is by no means necessarily true - it will be violated, for example, if radiative
cooling is not efficient, leading to a family of advection
dominated accretion flow \citep{Abramowicz:95, Narayan:95b} or adiabatic
inflow-outflow solutions \citep{Blandford:99}.    

Gas angular momentum transport on galaxy-wide scales is also rather
uncertain, even though is it believed that the black hole's local gas
reservoir may need to be replenished
during the black hole lifetime. While galaxy mergers and cosmic filamentary 
inflows have been convincingly advocated \citep{Hernquist:89, Barnes:91,
  Birnboim:03, Keres:05} as mechanisms of very efficient gas transport inwards 
to $\sim \,{\rm kpc}$ scales, the exact magnitude of these processes is
affected by star formation efficiency,
as well as by the nature of stellar and black hole 
feedback processes \citep[e.g.][]{Springel:05, Vandevoort:11, Dubois:13, Nelson:15}. On
sub-kpc scales, an array of physical mechanisms have been invoked 
to explain the successive gas angular
momentum transport. These include
global and local bar and spiral instabilities \citep{Roberts:79}, with the possibility of a
``bars-within-bars'' mechanism
\citep{Shlosman:89}, clumpy, turbulent discs
\citep{King:06,Krumholz:06, Hobbs:11} or discs supported by
radiation pressure \citep{Thompson:05}. While the
feasibility of some of these
mechanisms has been studied by means
of high resolution cosmological or isolated galaxy
simulations \citep[e.g.][]{Mayer:07, Levine:08,
  Hopkins:10, Emsellem:15}, gas
angular momentum transfer within
self-gravitating discs remains an
unsolved issue \citep{Goodman:03}. 

It is thus unsurprising that currently, most cosmological simulations estimate the growth of black holes by
use of a Bondi-Hoyle-Lyttleton approach \citep{Hoyle:39, Bondi:44, Bondi:52},
adopting the formula 
\begin{equation}
\dot{M} \,=\, \frac{4\uppi G^2M^2_\mathrm{BH}\rho_\infty}{(c^2_\infty + v^2_\infty)^{3/2}}\,,
\end{equation}
where $M_\mathrm{BH}$ is the mass of the black hole, $c_\infty$ and
$\rho_\infty$ are the sound speed and gas density at infinity and $v_\infty$
is the relative velocity between the
black hole and the gas at
infinity. Clearly, this does not take
into account the effect of gas 
angular momentum. Doing so is difficult - in galaxy formation simulations we
cannot directly resolve the relevant scales required to understand the
accretion flows in regions close to the
black hole (i.e. in the sub-kpc or sub-parsec
regime) and, as discussed above, even if we could we are fundamentally limited
by our understanding of the physical processes on these scales. What is
required to improve on the
Bondi-Hoyle-Lyttleton approach in
cosmological simulations is a
simple theoretical
framework that links how the accretion rate on small scales may be affected by
the mean angular momentum of the in-falling gas on large scales.

There have been several attempts to account for the effects of angular
momentum on the black hole accretion rate within galaxy formation
simulation. \citet{Power:11} use an accretion-disc particle technique, whereby
gas particles of sufficiently low angular momentum that their orbits would
cross a specified accretion radius are added to a particle that accounts for
the unresolved inner accretion disc. The black hole then accretes matter from
this disc on some viscous timescale. \citet{Rosas} make similar assumptions,
and calculate the net tangential velocity of accreting material and compare
this to the sound speed of the gas. For sufficiently high gas angular
momentum, the traditional Bondi rate is suppressed.

In this paper we adopt an alternative method, based on the work of
\citet{Krumholz} (for previous related works see also \citet{Abramowicz:81,
  Proga:03}) 
who parameterise the accretion rate on to a central object in 
terms of the vorticity of the gas on large scales. The effect of increased
vorticity is to suppress the accretion rate from the standard Bondi rate, as
gas with a larger impact parameter than the accretion radius of the black hole
circularises before it can cross the horizon and forms a disc. In the limit of
no vorticity, the prescription reduces to the spherically symmetric Bondi
accretion case.

We implement this vorticity-based accretion rate prescription in the moving
mesh code {\small 
  AREPO}. After validating our implementation on a number of stringent
tests, we investigate how this affect the accretion rate on to supermassive
black holes in
simulations of isolated galaxies, as well as of some simple galaxy binary
mergers. In doing so, we make use of the refinement technique as described by
\citet{Curtis:15}, which allows us to improve the resolution of the fluid flow
around black holes in our simulations. Accurately resolving the velocity
structure of the gas in this region is vital for successfully implementing the
vorticity prescription and we show how failing to do so leads to an
underestimate of the vorticity. Whilst this work focuses on galaxy-scale
simulations, we intend to present a method for use in fully cosmological
simulations.

This paper is organized as follows. In Section~\ref{Methodology} we outline
our methodology and the exact nature of our new implementation. In
Section~\ref{Analytic} we discuss analytical expectations of the
vorticity-based accretion rate prescription on the dynamical models of black
hole growth. We then present several numerical experiments in
Section~\ref{Validation} to validate our numerical implementation, before
showing the main results of our work in Section~\ref{Implications} where both
isolated and merging galaxy simulations are discussed. Finally,
Section~\ref{Conclusions} presents our conclusions and plans for future work.

\section{Methodology}\label{Methodology}
\subsection{Code}
\subsubsection{Basic setup}
We use the moving-mesh code {\small AREPO} code \citep{Springel:10} for all
simulations in this paper. {\small AREPO} employs the  {\small TreePM} approach
\citep{Springel:05} for handling gravitational interactions, and dark matter
is represented by a collisionless fluid of massive particles. The fluid is
modelled using a quasi-Lagrangian finite volume technique whereby the gas is
discretized by tessellating the computational domain with a Voronoi mesh. 

Star formation is carried out using the model of \citet{SF}. We stochastically
form star particles from cells of gas above a density threshold of
$\rho_\mathrm{sfr} = 0.18\mathrm{cm}^{-3}$, with a characteristic time-scale of
$t_\mathrm{sfr}=1.5\mathrm{Gyrs}$. Once formed, star particles only interact
gravitationally with other particles. The unresolved thermal and turbulent
processes that occur in the interstellar medium (ISM) are assumed to result in
a self-regulated  equilibrium described by an effective equation of state. We
adopt two different such equations: the most commonly used multiphase model
with a ``stiff'' equation of state ($q_{\rm EOS} = 1$ in the notation of
\citet{SF}), as well as a softer equation of state with $q_{\rm EOS} = 0.1$
, which interpolates between $q_{\rm EOS}  = 1$ and an
      isothermal medium of temperature $10^4$K, and which corresponds to a
      less pressurized and more clumpy medium. In addition to this, in some
      simulations we include metal line cooling as in \citet{Vogelsberger:13},
      where we assume that the gas has a solar abundance, since we are
      primarily interested in the qualitative range of any subsequent
      effects. The variations of the ISM model we consider here are aimed at
      addressing the possible impact of the gas thermodynamical state on the
      subsequent nature of the accretion flow and hence black hole growth.

\subsubsection{Black hole refinement scheme}
In order to improve the resolution of the simulation in the region of black
hole particles we employ the black hole refinement scheme of
\citet{Curtis:15}. This makes use of the ability of {\small AREPO} to split
and merge cells based on arbitrary criterion by forcing cells to decrease
linearly in radius as they approach the black hole. This means that for cells
a distance $d$ from a black hole particle have a cell radius $R$, where
\begin{align}
\frac{d}{R_\mathrm{ref}}\frac{(R^\mathrm{cell}_\mathrm{max} - R^\mathrm{cell}_\mathrm{min})}{c} +  \frac{R^\mathrm{cell}_\mathrm{min}}{c} &< R\,, \\
\frac{d}{R_\mathrm{ref}}(R^\mathrm{cell}_\mathrm{max} - R^\mathrm{cell}_\mathrm{min}) +  R^\mathrm{cell}_\mathrm{min} &> R\;.
\end{align}
Here $R_\mathrm{ref}$ is the scale over which refinement occurs, which we
typically set to be equal to the black hole smoothing length, $c$ is a
constant that we set to $2$, whilst $R^\mathrm{cell}_\mathrm{min}$ and
$R^\mathrm{cell}_\mathrm{max}$ determine the aggression of the
refinement. We typically set these to be equal to the Bondi radius and the
black hole smoothing length, respectively.

\subsection{Black hole model}
We adopt the model of \citet{Curtis:15}, based upon \citet{DiMatteo:05, Springel:05} to follow the evolution of black holes. We compare our vorticity-based prescription outlined below to that of the standard Bondi-like formula
\begin{equation}
\dot{M}_\mathrm{Bondi} = \frac{4\uppi G^2 M^2_\mathrm{BH} \rho_\infty}{c_\infty^{3}}\,,
\end{equation}
where we do not take into account the relative
velocity term in this work. Moreover in all simulations
we limit the accretion on to the black hole to the Eddington limit
\begin{equation}
\dot{M}_\mathrm{Edd} \equiv \frac{4\uppi GM_\mathrm{BH}\mathrm{m}_\mathrm{p}}{\epsilon_\mathrm{r}\sigma_\mathrm{T}c}\,,
\end{equation}
where $\sigma_T$ is the Thomson cross section, $m_p$ is the mass of a proton
and $\epsilon_\mathrm{r} = 0.1$ is the radiative efficiency.

In this paper, feedback is injected by adopting several different mechanisms,
as we are interested in the effect that this choice has on the magnitude of
the black hole growth suppression given by our vorticity implementation. This is driven by the fact that we expect that, at high sound speeds, the suppression will be greatly reduced. We proceed by coupling a fraction $\epsilon_\mathrm{f}$ of the luminosity to the gas, leading to an energy budget of
\begin{equation}
\Delta E_\mathrm{feed} = \epsilon_\mathrm{f}\epsilon_\mathrm{r}\Delta M_\mathrm{BH} c^2\,.
\end{equation} 
Here, $\epsilon_\mathrm{f} = 0.05$ is the feedback efficiency, which is set to reproduce the normalization of the locally inferred $M_\mathrm{BH} -
\sigma$ relation \citep{DiMatteo:05, Sijacki:07}.

We allow for feedback to occur both isotropically (pure thermal coupling) and
in a bipolar manner. In the latter case, we inject mass, energy and momentum
within a cone of opening  angle $\theta_\mathrm{out} =
\uppi/4$ which is aligned with the net angular momentum axis of the accreting
gas. In this case, the black hole is restricted to accrete from the cold disc
and we estimate the accretion rate using parameters derived from this
component. In bipolar simulations when a cold disc is very subdominant, we
proceed as in the isotropic case. For full details, see \citet{Curtis:15}.

\subsubsection{Vorticity implementation}

The benefits of the refinement method mean that in principle it is possible to account for the angular momentum of the gas in different ways \citep{Power:11,Rosas}. In this paper we implement the unified model of \citet{Krumholz}, which is very well suited to our computational approach. For completeness, we briefly summarize the method here. The authors consider an accreting point particle of mass $M$ within a velocity field of finite vorticity at infinity, which they parameterize as
\begin{equation}
\bm v_\infty = \omega_\star c_\infty \frac{y}{r_B}\,\hat{x}\,,
\end{equation}
where $r_B$ is the Bondi radius, $c_\infty$ is the sound speed of the gas at
infinity, and $\omega_\star$ is a dimensionless
`vorticity parameter', so called because the vorticity of the flow is
$-\omega_\star c_\infty/r_B\,\hat{z}$, where we have assumed a Cartesian coordinate system centred on the black hole. The task is then to analyse the
accretion as a function of $\omega_\star$.  For $\omega_\star = 1$, gas with impact parameter equal to the Bondi radius is travelling at the Keplerian velocity. This suggests that there is a transition region around $\omega_\star = 1$, which leads \citet{Krumholz} to consider three regimes of accretion:
\begin{enumerate}
\item Very small $\omega_\star$, where original Bondi accretion holds.
\item Small $\omega_\star$
\item Large $\omega_\star$
\end{enumerate}

\paragraph{Very Small $\omega_\star$} As the work attempts to build on the
Bondi formula, it is useful to consider first the cut-off for which the
angular momentum is too small to have an effect. \citet{Krumholz} argue that
if the gas only circularizes at a radius that is less than that of the
accreting object (in this case the Schwarzschild radius, $\mathrm{r}_\mathrm{s}$, as we consider
non-spinning black holes) then the material will be accreted on mostly radial orbits. This implies a condition for when the original Bondi rate is appropriate
\begin{equation}
r_\mathrm{circ} = \omega_\star^2y^4/\mathrm{r}_\mathrm{B}^3 < \mathrm{r}_\mathrm{s}\,,
\end{equation}
which implies
\begin{equation}
\omega_\star < \omega_\mathrm{crit} \equiv \sqrt{\mathrm{r}_\mathrm{s}/\mathrm{r}_\mathrm{B}}\,,
\end{equation}
as the condition for the original Bondi prescription to hold.
\label{very_small_omega}

\paragraph{Small $\omega_\star \ll 1$, $\omega_\star > \omega_\mathrm{crit}$} \label{smallom}
Above the critical vorticity derived in the previous section, the gas will circularise before it reaches the accretor. In these cases, the analysis of \citet{Krumholz} suggests that the resulting torus that forms will extend to $r_B$ with a scale height of $\sim r_B$, thus blocking accretion in the range $\uppi/4<\theta<3\uppi/4$, i.e. in the plane of the gaseous disc. This has the effect of decreasing the accretion rate to $\dot{M} \sim 0.3\dot{M}_\mathrm{Bondi}$.

\paragraph{Large $\omega_\star \ge 1$} For the final case, that of a high vorticity, \citet{Krumholz} argue that accreted material will lie on streamlines that bend sharply, shocking the gas and allowing it to lose energy and fall on to the central particle. This will occur on streamlines that have a larger peak escape velocity than the velocity of the gas at infinity. The accretion rate can then be deduced by summing the gas entering an area $A$ where the condition is satisfied, that is
\begin{equation}
\dot{M} \approx \int_A\rho_\infty\sqrt{v_\infty(y,z)^2+c^2_\infty}\,\mathrm{d}y\,\mathrm{d}z\,,
\end{equation}
which, when calculated, gives
\begin{equation}\label{approxmdot}
\dot{M}(\omega_\star) \approx 4\uppi r_\mathrm{B}^2\rho_\infty c_\infty f(\omega_\star)\,,
\end{equation}
where $f(\omega_\star)$ is a numerical factor that is approximated in the limit $\omega_\star\to\infty$ by
\begin{equation}
f(\omega_\star) \approx \frac{2}{3\uppi\omega_\star}\mathrm{ln}(16\omega_\star)\,.
\end{equation}

We implement this scheme into the {\small AREPO} code. We calculate the vorticity parameter as
\begin{equation}
\omega_\star=\lvert\bm{\omega}\rvert\frac{\mathrm{r}_\mathrm{B}}{c_\infty}\,,
\end{equation}
by taking an exponential spline kernel-weighted \citep{Monaghan:85} mean of the vorticities of the
neighbouring cells to the black hole\footnote{Note that we have explored
  different choices of how many neighbouring gas cells to consider (e.g.
  within $0.1$, $0.3$, $0.5$ and $1$ of the black hole's smoothing
  length). While the exact value of the vorticity estimate may in some cases be
  sensitive to this choice, we found that our estimate of the accretion rate for all our results using
  super-Lagrangian refinement were robust and largely unchanged. This is further supported by our analytic
  estimates 
  presented in Section~\ref{Analytic}.}. We then calculate the accretion rate whenever $\omega_\star > \omega_\mathrm{crit}$ using the following expression
\begin{equation}\label{eqn:vort}
\dot{M}_{\rm Vort} = \frac{4\uppi G^2M_{\rm BH}^2\rho_\infty}{c_\infty^3} \times \begin{cases} 0.34 &\mbox{ } \omega_\star < 0.1 \\
0.34 \cdot \frac{2}{3\uppi\omega_\star}\mathrm{ln}(16\omega_\star) & \mbox{ } \omega_\star > 0.1\,. \end{cases} 
\end{equation}
since the analytic expression is well fit by a piecewise approximation
  for $\omega_\star$ greater and less than $0.1$. The factor of $0.34$
  accounts for the same phenomenon detailed in Section 
\ref{smallom}, selected to match the value of $0.3\dot{M}_\mathrm{Bondi}$ for
$\omega_\mathrm{crit} < \omega_\star \ll 1$. It differs from exactly $0.3$
because of a slight difference between equation (\ref{approxmdot}) and the
standard Bondi rate.

\subsection{Initial Conditions}
\label{ICs}
In order to understand the implications of our implementation, we have carried
out a series of simulations of an isolated disc galaxy. This simulation setup
is particularly well suited to our topic of inquiry as, on large scales, the
majority of gas will be rotationally supported and thus the effect of gas
angular momentum on the black hole accretion rate may be significant. In fact,
the observed luminosities of local AGN harboured within spiral galaxies are
either rather low or modest, and in the latter case often fall within the
category of Seyferts, suggesting limited fuelling rates possibly driven by bar
instabilities \citep[see e.g.][]{Knapen:00, Laine:02, Kewley:06}.

Our initial conditions match those of \citet{Curtis:15} (for original
implementation of the model and further details see \citet{Springel:05}). These consist of an isolated galactic disc in hydrostatic equilibrium, situated within a dark matter halo described by a \citet{Hernquist:90} density profile
\begin{equation}
\rho_\mathrm{dm} = \frac{M_\mathrm{dm}a}{2\uppi r(r+a)^3}\,,
\end{equation}
where $M_\mathrm{dm}$ is the total dark matter mass and $a$ is a scaling
parameter. We generate a self-consistent gaseous disc distribution in equilibrium and
potential by an iterative process, based on an exponential surface density
profile, with the azimuthal velocity set to
\begin{equation}\label{eqn:velocity}
v_{\rm \phi,gas}^2 = R\left(\frac{\partial \Phi}{\partial R} +
\frac{1}{\rho_{\rm gas}}\frac{\partial P}{\partial R}\right)\,,
\end{equation}
and the remaining velocity components $v_R = v_z = 0$. We also include a
centrifugally supported stellar disc that also follows an exponential surface
density profile. Our parameters are chosen to represent a simplified Milky
Way-like galaxy, but note that we do not attempt to explicitly model
Sagittarius A* in any way. 

In the last part of this paper we contrast our isolated disc simulations with
simulations of binary galaxy mergers. We consider only equal mass mergers on
prograde parabolic orbits and initial galaxy properties that are identical to our
isolated galaxy setup. These simulations serve to explore the difference in 
black hole growth reduction due to the vorticity-based accretion rate prescription
in the case where large scale gravitational torques affect the central gas density
and angular momentum distribution significantly.  

In Table~\ref{Table1} we list all simulations of isolated and merging
galaxies used in this work, where we summarize the initial number of gas
cells, type of refinement method, black hole accretion and feedback
prescriptions and any other variations in the models, which relate to black hole
seed mass, type of radiative cooling and the effective equation of state for
the ISM.  

\begin{table*}
\bc
\begin{tabular}{ccccccccc}
\hline\hline
Type & $N_\mathrm{gas}$ & Refinement & Accretion & Feedback & Additions\\ \hline
Isolated & $10^5$ & No & Bondi & No & -\\
Isolated & $10^5$ & Yes & Bondi & No & -\\

Isolated & $10^5$ & No & Bondi & Thermal & -\\
Isolated & $10^6$ & No & Bondi & Thermal & -\\
Isolated & $10^7$ & No & Bondi & Thermal & -\\

Isolated & $10^5$ & Yes & Bondi & Thermal &- \\
Isolated & $10^6$ & Yes & Bondi & Thermal &- \\
Isolated & $10^7$ & Yes & Bondi & Thermal &- \\

Isolated & $10^5$ & Yes & Bondi & No & Metal line cooling\\
Isolated & $10^5$ & Yes & Vorticity & No & Metal line cooling\\

Isolated & $10^5$ & Yes & Bondi & No & sEOS\\
Isolated & $10^5$ & Yes & Vorticity & No & sEOS\\

Isolated & $10^5$ & Yes & Bondi & No & $M_0 = 10^7 M_\odot$\\
Isolated & $10^5$ & Yes & Vorticity & No & $M_0 = 10^7 M_\odot$\\

Isolated & $10^5$ & No & Vorticity & Thermal & - \\
Isolated & $10^5$ & Yes & Vorticity & Thermal & - \\

Isolated & $10^5$ & Yes & Bondi & Bipolar & sEOS\\
Isolated & $10^5$ & Yes & Vorticity & Bipolar & sEOS\\

Merger & $10^5$ & Yes & Bondi & No & Metal line cooling\\
Merger & $10^5$ & Yes & Vorticity & No & Metal line cooling\\
Merger & $10^6$ & Yes & Bondi & No & Metal line cooling\\
Merger & $10^6$ & Yes & Vorticity & No & Metal line cooling\\

\hline\hline
\end{tabular}
\caption{Simulation details for isolated and merger galaxy models, as
  indicated in the first column. The second column lists the initial number of gas
  cells employed (note that due to super-Lagrangian refinement this number can
increase significantly during the simulated time-span). The third column indicates
for which models we use our super-Lagrangian refinement around black
holes (Yes) and for which we only use the standard (de)/-refinement present in
\small{AREPO} (No). The fourth and fifth columns indicate our adopted model for
black hole accretion and feedback, while in the sixth column we list any other
variations considered, such as the metal line cooling in addition to the
primordial cooling (which is always switched-on), a softer equation of state
(sEOS) for the ISM and a larger black hole seed mass of $10^7\, {\rm
  M_\odot}$, instead of our default choice of $3.8 \times 10^5\, {\rm
  M_\odot}$.}\label{Table1}
\ec
\end{table*}
 
\section{Analytic Results: Expected Suppression of Black Hole Growth} \label{Analytic}

The black hole accretion rate, for both the vorticity and the standard Bondi
rate prescription, is highly sensitive to the gas sound speed, gas density and the
black hole mass. Probing the full parameter space with numerical simulations
is non-viable, both because of the non linear nature of the evolution but also
because of the significant computational time to carry out an individual
simulation. To tackle this, we here explore the effects of changing the gas
and black hole properties in the context of our isolated galaxy setup (see
Section \ref{ICs}) and we subsequently explore part of the parameter space
with fully self-consistent simulations. 

It should be noted here that in both observations and simulations the exact
nature of the gas close to the black hole's Bondi radius is poorly
understood \citep{Frank:02, Goodman:03}. While a number of recent studies
constrain gas properties close to the Bondi radius in some systems, such as
Sagittarius A* \citep{Yuan:03, Xu:06}, M87 \citep{Russell:15} or
typically in LINERs and Seyferts through mega-masers \citep[for a review
  see][]{Lo:05}, we do not yet have a consensus on the density, temperature and
velocity distribution of the gas. Thus for the following analysis, we make a very simplifying assumption
 and consider that the gas velocity field matches the initial (numerically derived) conditions down to the Bondi radius (see equation \ref{eqn:velocity}), although we also investigate the effect of assuming a field with a higher vorticity. The vorticity profile scales approximately as $\bm{\omega} \sim (R^{0.5})\vec{\hat{z}}$ with radius.

\label{analytic_distributions}

\subsection{Expected suppression factor}

\begin{figure*}
\centering
\includegraphics[width=0.49\textwidth]{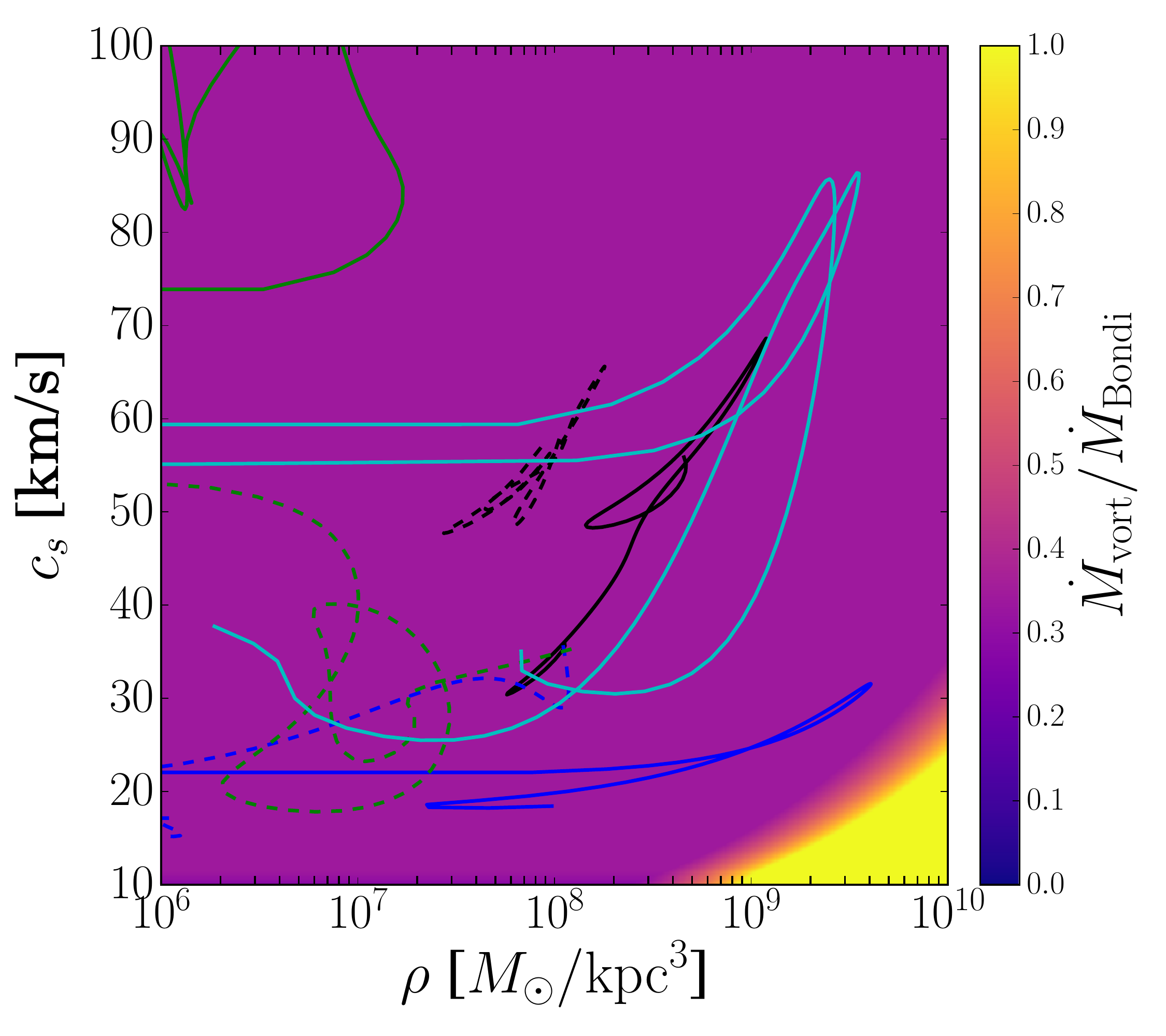}
\includegraphics[width=0.49\textwidth]{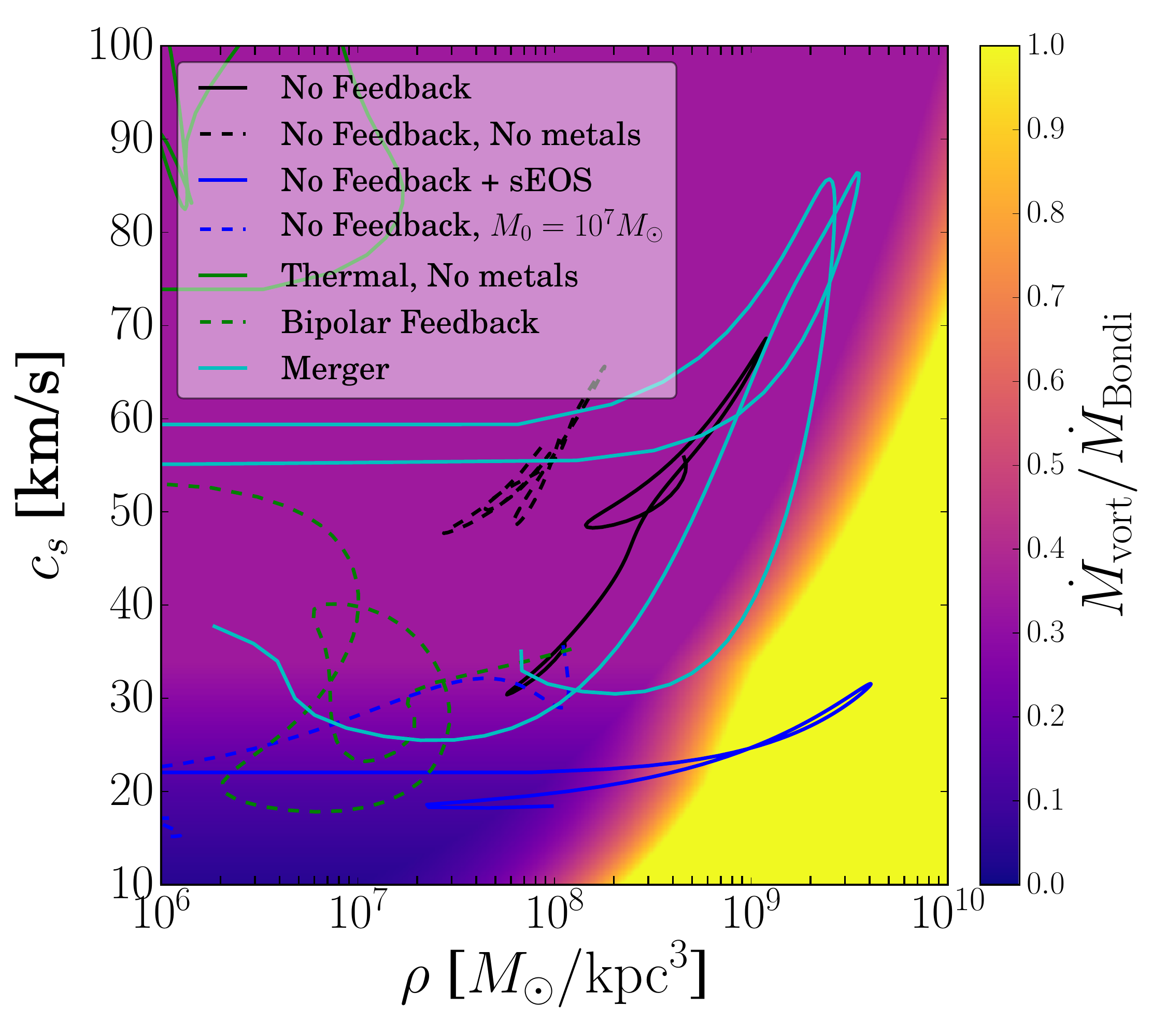}
\caption{Coloured 2D histograms showing the ratio
  $\min(\dot{M}_\mathrm{Vort},\dot{M}_\mathrm{Edd}) / 
  \min(\dot{M}_\mathrm{Bondi},\dot{M}_\mathrm{Edd})$ for two fixed black hole
  masses of $3.8 \times 10^5 M_\odot$ (left-hand panel) and $10^7 M_\odot$
  (right-hand panel) as a function of gas density and sound speed (see equation \ref{eqn:vort}), assuming a
  given gas vorticity profile. Overplotted, we show the tracks from several of our simulation models,
indicating the typical regimes simulated black holes are in (but note that
simulated black hole masses grow with time along the shown tracks). In the
bottom left region of the parameter space, for high densities and low sound
speeds (coloured yellow), the Eddington rate is the dominant limitation of
the accretion rate, as the Bondi rate is very high. However, unless the sound
speed is sufficiently small to allow for a large value of $\omega_\star$, the
suppressing effect of the vorticity is limited to 0.34. As such, the resulting
suppression is sensitive to the presence and type of feedback injected, as
well as the ISM physics. There is also a dependence on the mass of the central
black hole, as this affects the Bondi radius. Higher mass black holes present
larger regions of the parameter space with high suppression, but also a larger
region in which the black hole growth is Eddington limited.}
\label{suppression} 
\end{figure*}

In Figure~\ref{suppression}, coloured 2D histograms show the ratio of the
vorticity-suppressed black 
hole accretion rate to the standard Bondi rate as a function of gas density
and sound speed, for two fixed black hole masses of $3.8 \times 10^5
{\rm 
  M_\odot}$ (our default seed mass) and $10^7 {\rm M_\odot}$, in the left-hand
and right-hand panels, respectively. Here, we have included the effect of
limiting the black hole growth by capping the accretion rate at the Eddington
limit, hence the suppression is
$\min(\dot{M}_\mathrm{Vort},\dot{M}_\mathrm{Edd}) /
\min(\dot{M}_\mathrm{Bondi},\dot{M}_\mathrm{Edd})$. This means that for
sufficiently low sound speeds and high densities, the effective suppression is
zero because the growth is Eddington limited, rather than being constrained by
the angular momentum barrier (see yellow regions on the panels). At the other
end of the spectrum, we note that for the majority of the parameter space the
suppression ratio is at most $0.34$. This is because, even for large black
hole masses, the condition that the effective impact parameter be less than
the accretion radius of the black hole (see Section~\ref{very_small_omega}) is
not satisfied given our assumption on the velocity field of the gas. Indeed,
we shall see later that this condition is particularly difficult to satisfy,
even when large scale torques during a major merger drive a substantial mass
of gas to the central region of the galaxy. For the black hole with a
  higher mass there is a small region of the parameter space for which the
  suppression is especially high (see dark blue region on the right-hand
  panel) - this is driven by the increase in the Bondi radius of the black
  hole. Increasing the vorticity over that of our model by a factor of 10 has
  a similar effect, increasing the expected suppression for sound speeds less
  than $20$ km/s and slightly shrinking the region which is Eddington
  limited.

On both plots, for context, we also show a selection of tracks from our
simulations, to indicate the typical properties of the gas surrounding the
black hole. Note however that in simulations black hole mass is not constant
and thus to provide some guidance on the magnitude of the effect we over plot
the same tracks on both panels. The assumptions we make about black hole
feedback, as well as the cooling physics of the surrounding gas and the ISM
clumpiness, can have significant consequences for the suppression we
expect. The simulations with no feedback show a tight relationship between
sound speed and density. This is not surprising, as we assume that gas that is
sufficiently dense to be star-forming follows an effective equation of state
\citet{SF}, which has the effect of forcing the gas to higher sound speeds at
larger densities. We can vary this by using a softer equation of state for the
ISM - the track for this simulation (dark blue, solid) reaches much lower
sound speeds.  

On the other hand, thermal feedback distributed isotropically (green, solid)
has the effect 
of dramatically heating the surrounding gas - especially that which is closest
to the black hole. Whilst this itself has the effect of reducing the accretion
rate, it means that such simulations are never in the regime that allows for
strong angular momentum-driven suppression according to our model. If we
restrict the feedback to a bipolar cone (green, dashed) - explicitly
decoupling the heated gas from a cold, accreting gas disc, we can reach
regimes where suppression is significant. 

This strong dependence on the temperature of the gas is unsurprising when we note that the amount that we suppress the standard Bondi rate is proportional to $\omega_\star$, where:
\begin{equation}
\omega_\star \propto \frac{\lvert\bm{\omega}\rvert}{c_s^3}\,.
\end{equation}
Whilst some of the viscous processes that transport angular momentum might be expected to increase with increased thermal motions of the gas, we are fundamentally limited by of our treatment of the ISM. Improving our modelling of the gas physics to self consistently resolve a realistic multiphase ISM is the subject of much ongoing research - in the future it will certainly be interesting to see how this will affect our understanding of the gas properties around the black hole.

\subsection{Expected delay in black hole growth}
\label{delay_section}
\begin{figure*}
\centering
\includegraphics[width=0.49\textwidth]{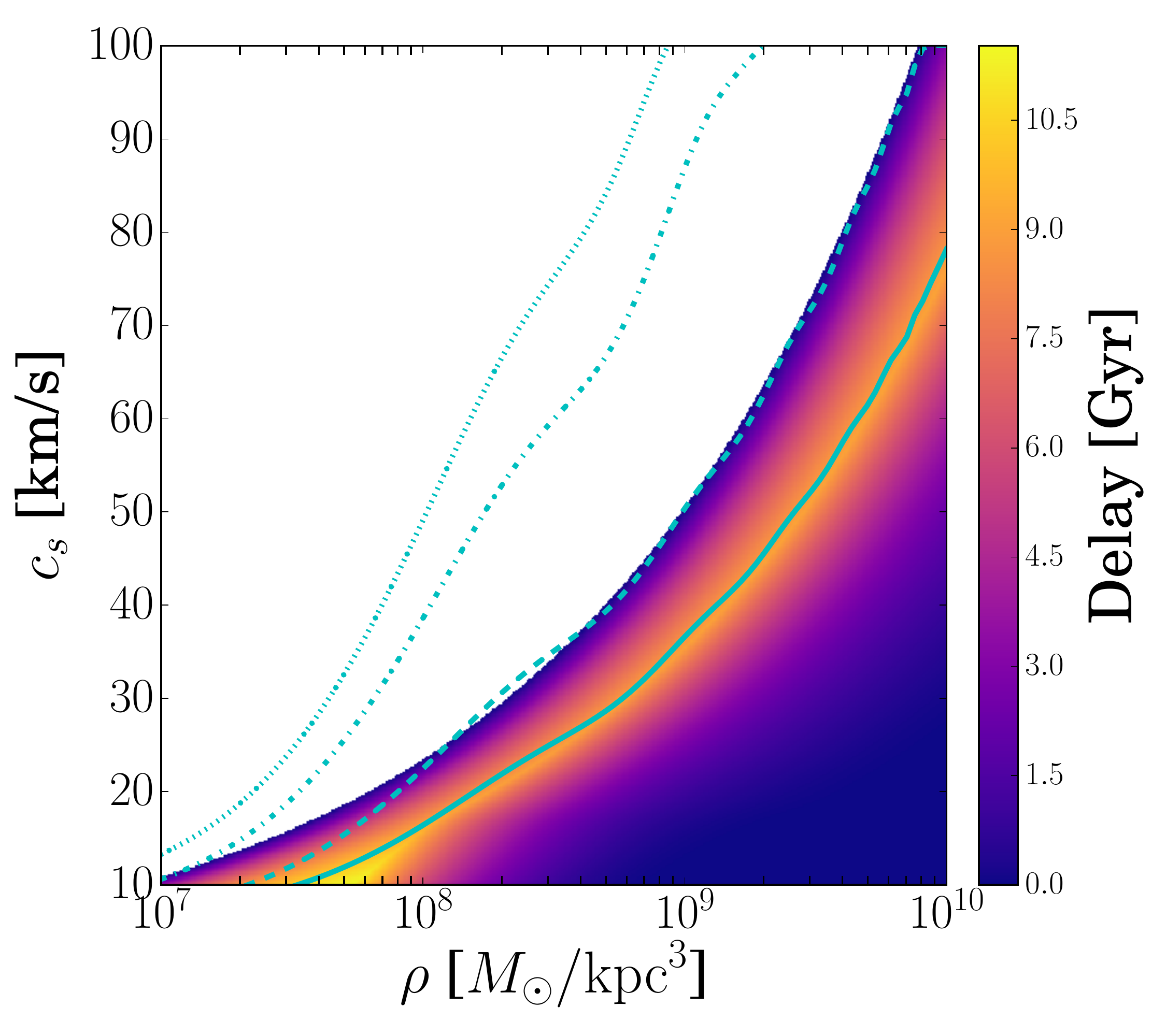}
\includegraphics[width=0.49\textwidth]{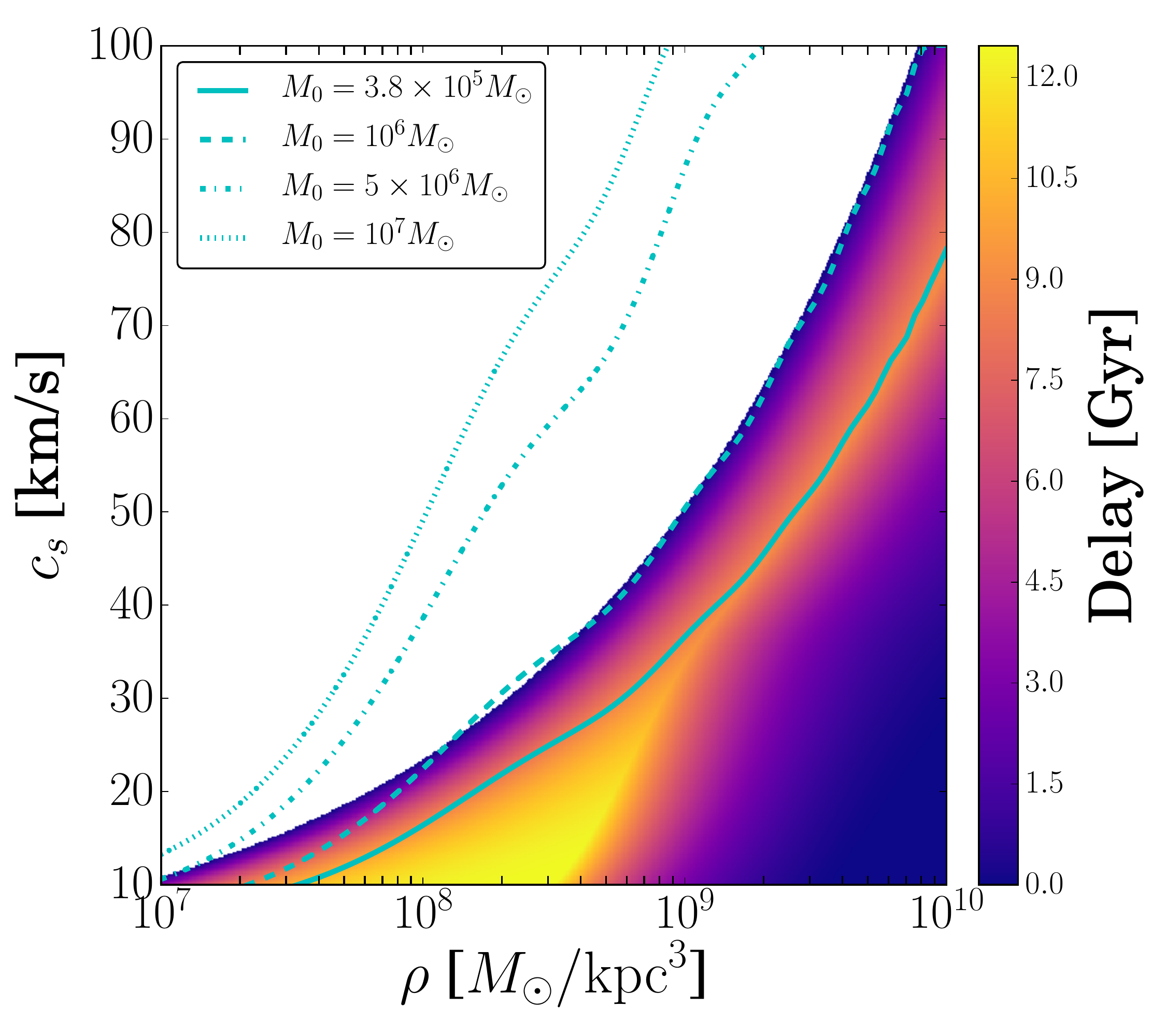}
\caption{2D coloured histograms showing the effective delay in Gyr for a $3.8
  \times 10^5 M_\odot$ black hole to increase in mass by an order of magnitude
  for our default gas vorticity profile (left-hand panel) and assuming a
  vorticity that is an order of magnitude larger than our initial conditions
  (right-hand panel). We plot the peak of this distribution for a series of
  different starting black hole masses with lines of different styles, as
  indicated on the legend. At high sound speeds, and low densities, the black
  hole growth is insignificant. Conversely, for large densities and low sound
  speeds, the Bondi rate is sufficiently high that the black holes grow very
  fast, regardless of the suppressing effect of the vorticity prescription. In
  between there is a clearly defined region of the parameter space (yellow to
  pink shades) where the suppressing effect has a significant impact, causing
  delays in black hole growth comparable with the Hubble time. For higher
  starting black hole masses lower densities and higher sound speeds lead to
  similar delays. In practice, whether this occurs in simulations depends on
  the evolution of the surrounding gas, which is highly dependent on both the
  feedback routine and the ISM model chosen.} 
\label{delay} 
\end{figure*}

Because of the non-linear nature of black hole growth within simulations,
looking at the ratio of the vorticity-based accretion rate to the standard
Bondi rate alone can be a misleading guide to the results of simulations. This
is principally because of two effects. Firstly, the Bondi rate itself can vary
dramatically with density and (in particular) sound speed of the ambient
medium. As such, if the black hole is growing either very fast or very slow,
then the final black hole mass will be dominated by the changes in the Bondi
rate itself, and the black hole growing according to the vorticity
prescription will undergo similar growth with a slight delay. Secondly, any
early growth in black holes is compounded at later times, because of the $M^2$
dependence of the Bondi rate. This is a problem that will be particularly
acute in simulations without strong self-regulation, which is the case for a
variety of different models of feedback and, of course, for simulations
without strong black hole feedback at all.

For this reason, we have carried out a series of simplified time integrations
of black hole growth. We make several basic assumptions: in particular we do
not include the effects of black hole feedback and we assume that the
distribution of gas properties as a function of radius remains the same over
time. We start the black holes at an initial seed mass and at each time step
we calculate the updated accretion rate based on sampling the relevant gas
properties at the Bondi radius (which grows with time), and continue the
integrations for a maximum of 14 Gyr. We will see that whilst the results from
simulated galaxies differ from these predictions in certain key ways, we are
able to reproduce the general effects of using the vorticity prescription over
a wide range of parameters.

In the left-hand panel of Figure~\ref{delay}, 2D coloured histogram shows the
delay in Gyr for a $3.8 \times 10^5 M_\odot$ black hole to increase its mass
by an order of magnitude, as a function of gas density and sound speed (and
keeping our default choice of gas vorticity distribution). The distribution is
clearly peaked, with a well defined maximum delay region, which we have over
plotted for different values of the initial black hole mass. The parameter
space for each is split into separate regions: in the top left (white region), 
for high sound speeds and low densities, the black hole is unable to grow by
any significant fraction. The opposite occurs for high densities and low sound
speeds (blue region) - the black hole grows fast regardless of
suppression. There is, however, a tight regime in which the delay in black hole
growth can be particularly significant (yellow-pink region) and comparable to
the Hubble time.  

On the right-hand panel of Figure~\ref{delay}, we show an analogous plot but
for an increased gas vorticity, which is now 10 times higher at all radii. For
a large region of parameter space, the resulting delay in black hole growth
remains similar or slightly larger. The region in which we expect the most
significant suppression increases in size, as the magnitude of the suppressing
effect increases, but the region of the parameter space in which we expect
significant delay is still relatively small. Finally, we note that for larger
initial black hole masses lower gas densities and higher sound speeds lead to
a similar effective delay. 

\section{Simulation Results: Validation of implementation} \label{Validation}

Whilst all simulations that attempt to implement sub-grid prescriptions face
difficulty in resolving the gas properties around the black hole, accurately
resolving a higher order quantity such as the velocity field is especially
difficult. In this section we detail our numerical experiments to validate our
implementation of the vorticity prescription within {\small AREPO}, whilst
indicating limitations with the current approach. Whilst much of our
discussion here is specific to the \citet{Krumholz} model, we note that our
investigations raise points that are important for alternative
approaches of angular momentum estimation in black hole accretion modelling. 

\subsection{Keplerian Disc}

\begin{figure*}
\centering
\includegraphics[width=0.49\textwidth]{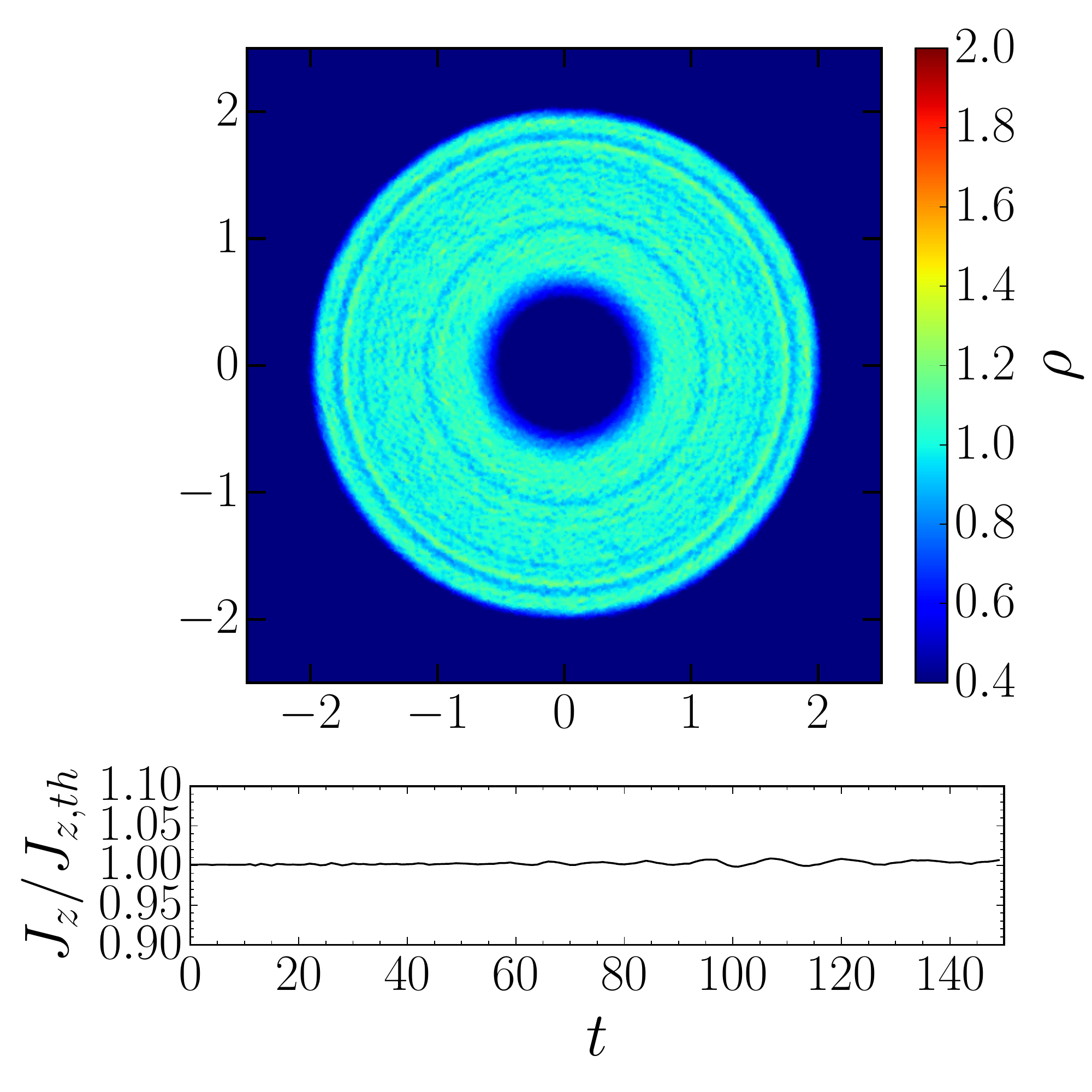}
\includegraphics[width=0.49\textwidth]{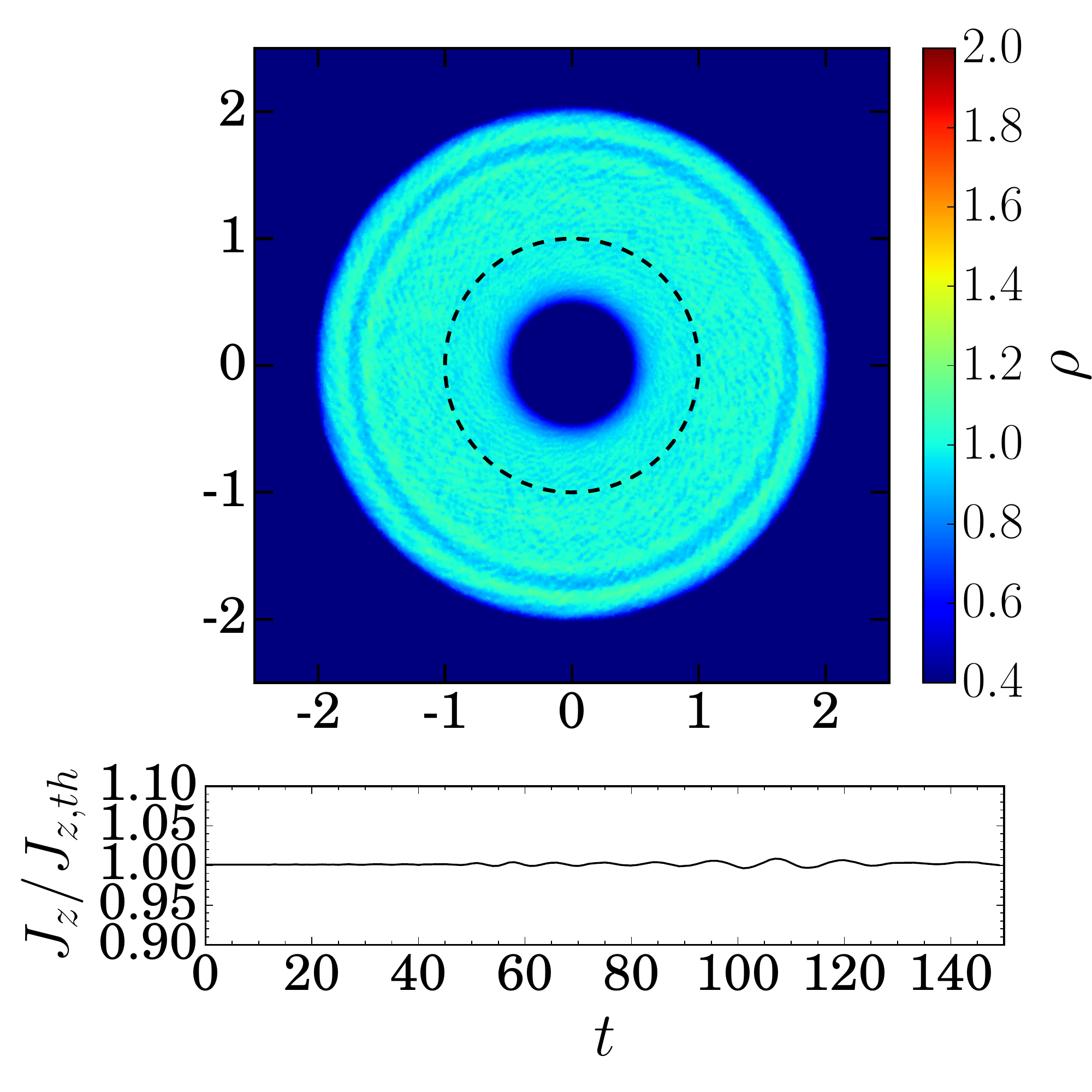}
\caption{In the left-hand panel we show the density map at $t=150$ of a 2D
  simulation of a cold Keplerian disc for a simulation with no cell refinement
  or de-refinement (note that the units used are arbitrary). In the right-hand panel we show an analogous plot, but for
  a simulation using our black hole refinement scheme, and the region within
  which we impose super-Lagrangian refinement is marked with a dashed
  circle. In both cases, we also show  the ratio of the total angular momentum
  in the simulation to the initial theoretical value (bottom row). The
  simulations show very good conservation properties for the entire simulated
  time-span. In particular, the aggressive cell refinement and de-refinement
  do not cause any significant differences to the angular momentum
  conservation of the code.} 
\label{kep_disc} 
\end{figure*}

In order to test the conservation properties of our refinement scheme, we
simulate a cold, pressure free Keplerian disc. This problem, which holds
modern hydrodynamics codes to a particularly stringent standard, is one that
has been used to demonstrate the effects of systematic errors in angular
momentum conservation \citep{Cullen:10, Hopkins:15, Pakmor:16}. In its latest
version, which in particular adopts improvements to the gradient estimation
and time integration, our moving mesh code {\small AREPO} shows very good
conservation properties - for full details see \citet{Pakmor:16}. Here, we are
interested in the effects that our super-Lagrangian refinement scheme has on
the gas angular momentum within the refinement region.

We adopt the same initial setup as \citet{Pakmor:16}. In particular, we
initialize a disc spanning a radial range of $0.5 < r < 2.0$ with gas density
$\rho = 1.0$, and gas velocities $v_x=-yr^{-3/2}$, $v_y=xr^{-3/2}$. Here, the units used are arbitrary. Outside of
this region, we set $\rho = 10^{-5}$ and $v=0$. In both regions we set
the internal energy to $u=2.5\time 10^{-5} \gamma/\rho$, where we set the
adiabatic index $\gamma = 5/3$. Our simulation uses a constant external
gravitational acceleration of 
\begin{equation}
g = -\frac{r}{r(r^2 + \epsilon^2)}\,,
\end{equation}
where $\epsilon = 0.25$ for $r<0.25$ and is zero otherwise. Since in this case
there is no black hole particle, in the simulation with super-Lagrangian
refinement we set the refinement region to be
$R_\mathrm{ref}\, =\,1.0$ and the maximum cell radius
$R_\mathrm{max}^\mathrm{cell} = 10^{-2}$, which represents the size required
for a smooth transition at the refinement boundary. We have tested different
values for $R_\mathrm{min}^\mathrm{cell}$ - since the disc in this case does
not extend to the centre of the simulation we find that our results are
relatively insensitive to the exact value. For the plots here we show
our simulation for which $R_\mathrm{min}^\mathrm{cell} = 10^{-4}$. 

Figure~\ref{kep_disc} illustrates the results for simulations both with
(right-hand 
panel) and without (left-hand panel) our super-Lagrangian refinement
scheme. We show the density across the disc at $t=150$, and note the good
agreement between the two simulations, which both show that the disc has
remained stable for the entire duration of the simulation. In particular, the
simulation with refinement shows somewhat more stable inner edge of the cold
disc. The ratio of the total angular momentum of the disc to the initial
theoretical value is plotted in the bottom row. Both simulations show very
good conservation properties throughout, indicating that the cell refinement
and de-refinement have not introduced neither significant angular momentum
transport nor conservation issues. This is reassuring as it demonstrates that
the estimation of gas angular momentum, and hence vorticity, does not suffer
from spurious numerical artefacts due to our super-Lagrangian refinement.

\subsection{Detecting spurious vorticity}

\begin{figure*}
 \centering
 \includegraphics[width=0.45\textwidth]{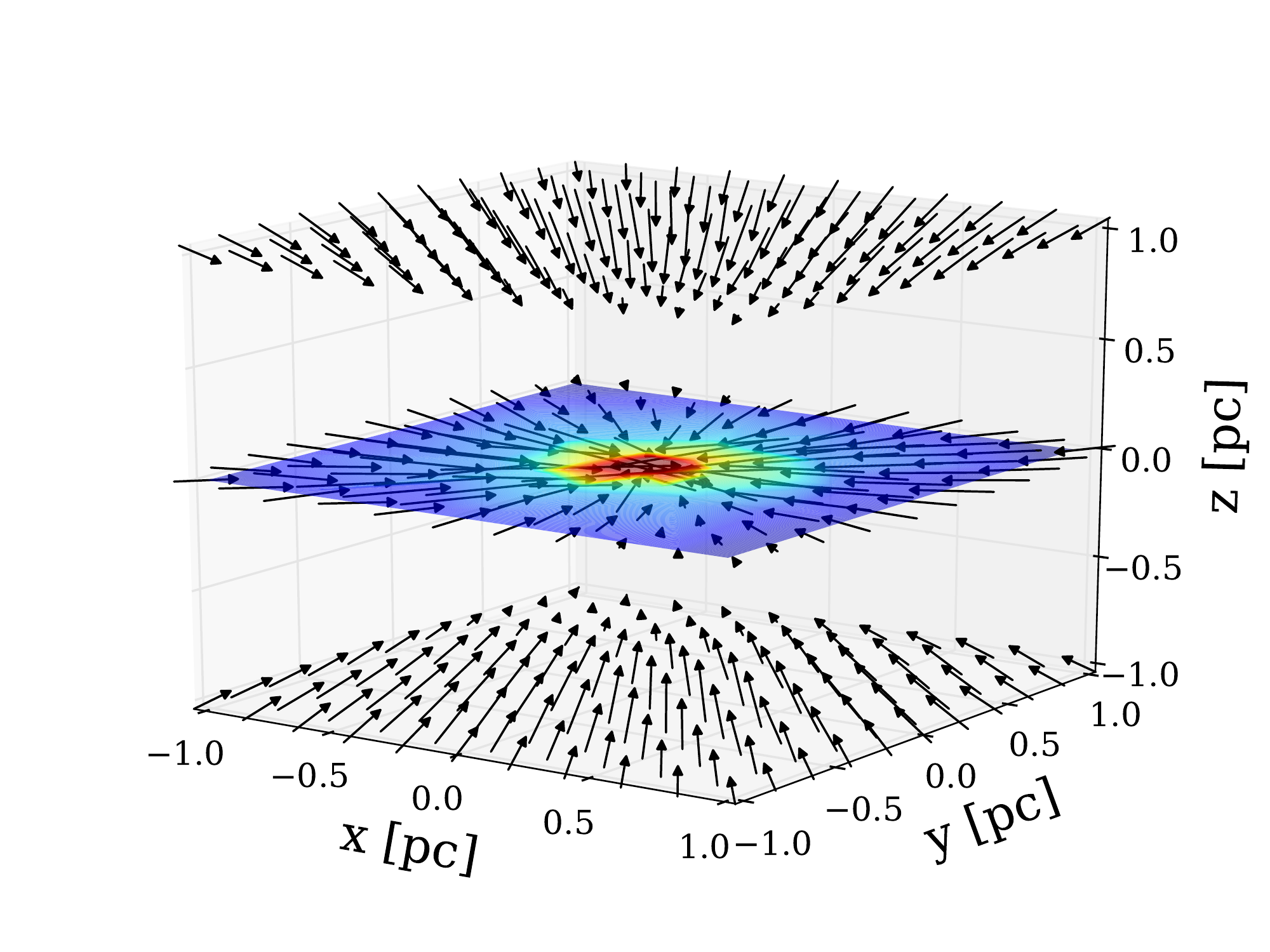}
 \includegraphics[width=0.45\textwidth]{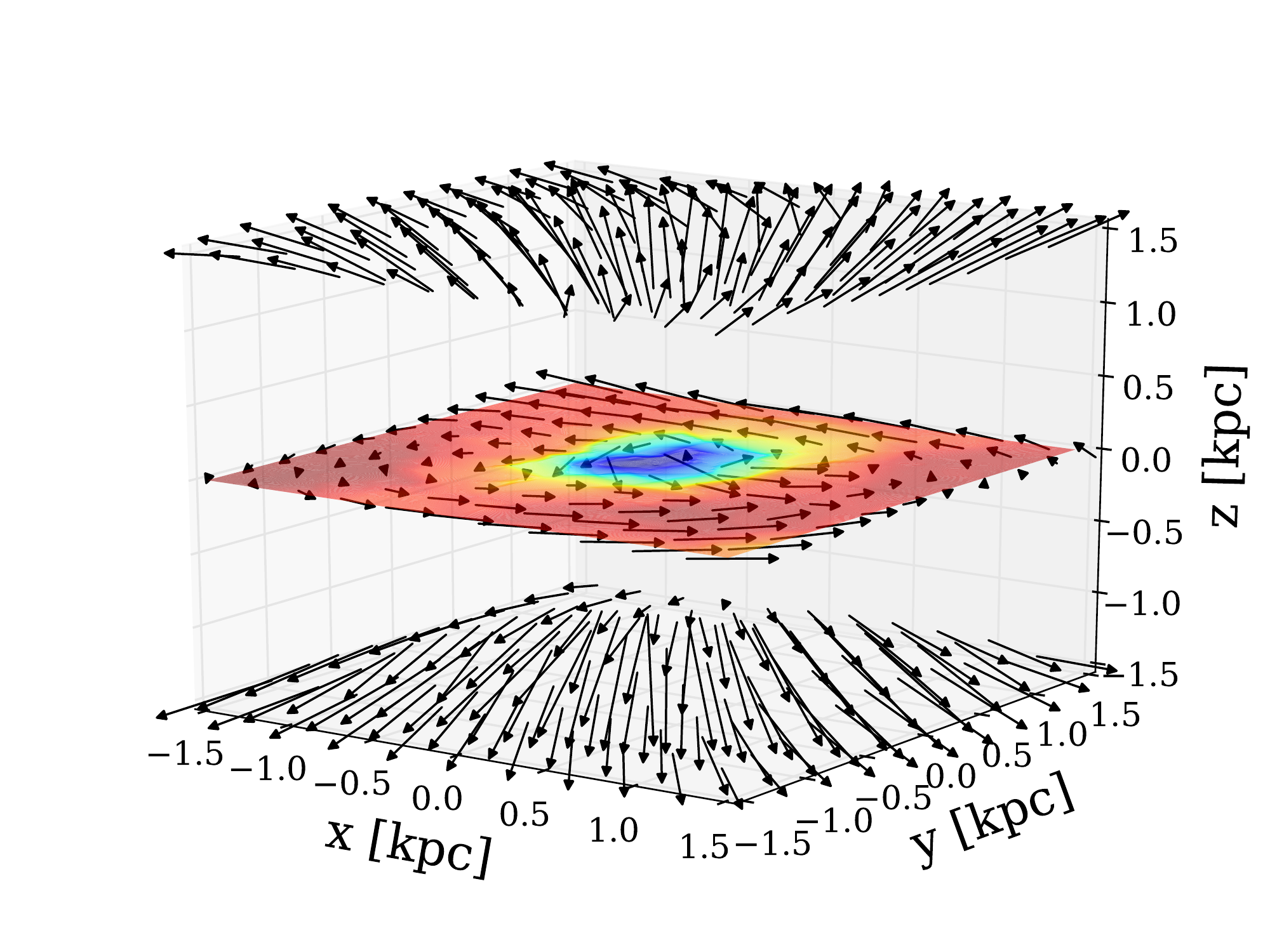}\\
 \includegraphics[width=0.45\textwidth]{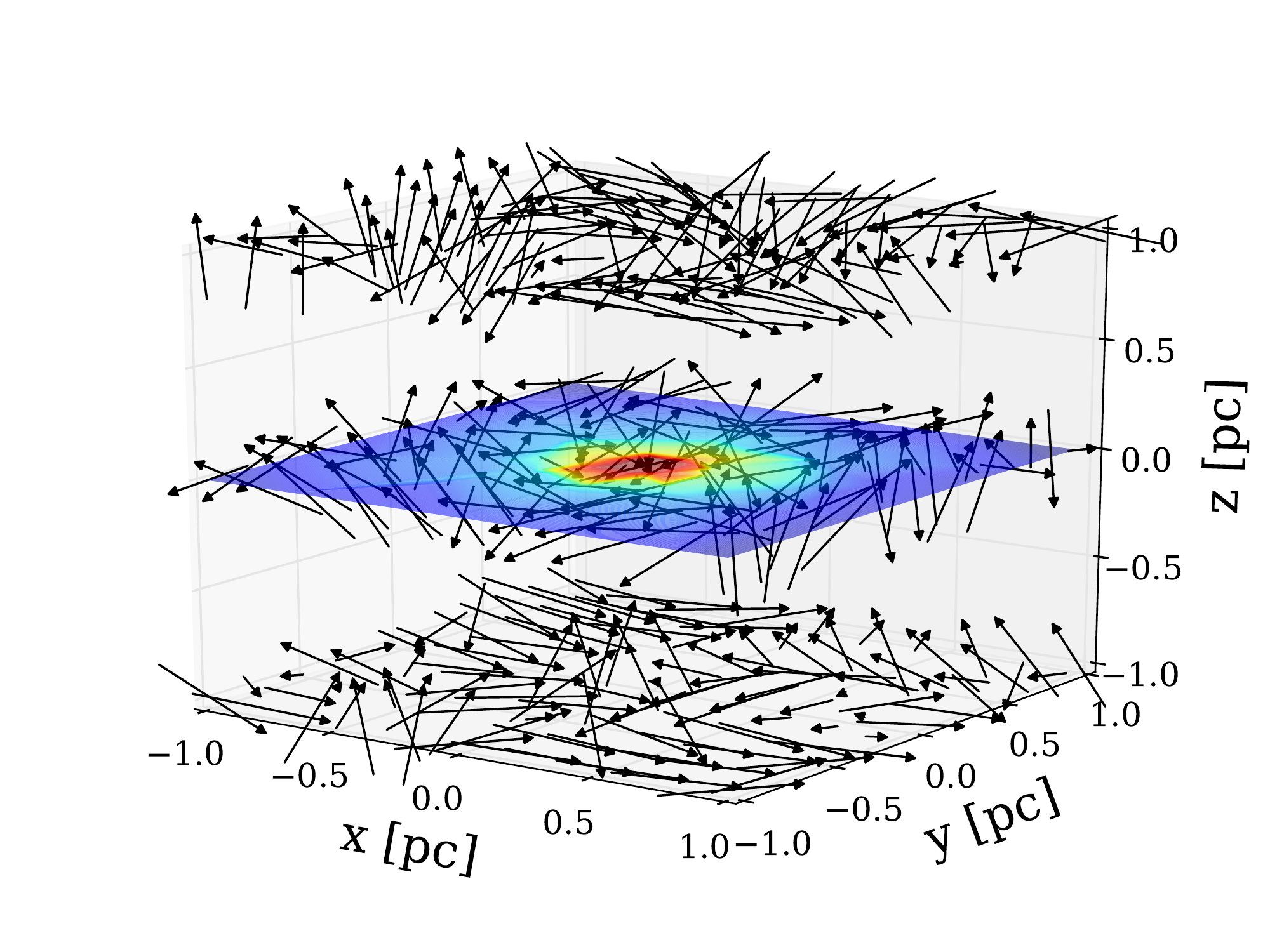}
 \includegraphics[width=0.45\textwidth]{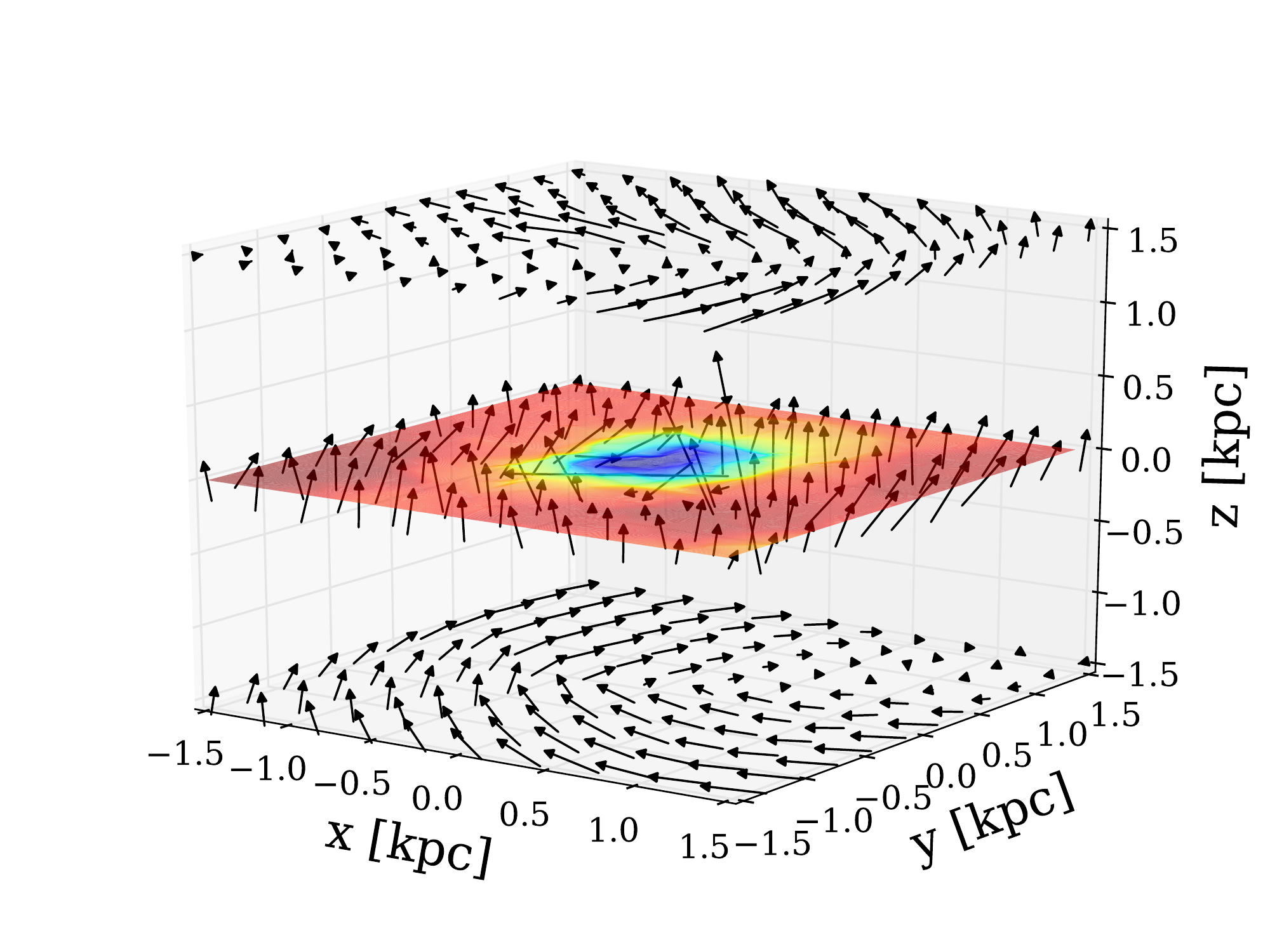}\\
 \includegraphics[width=0.45\textwidth]{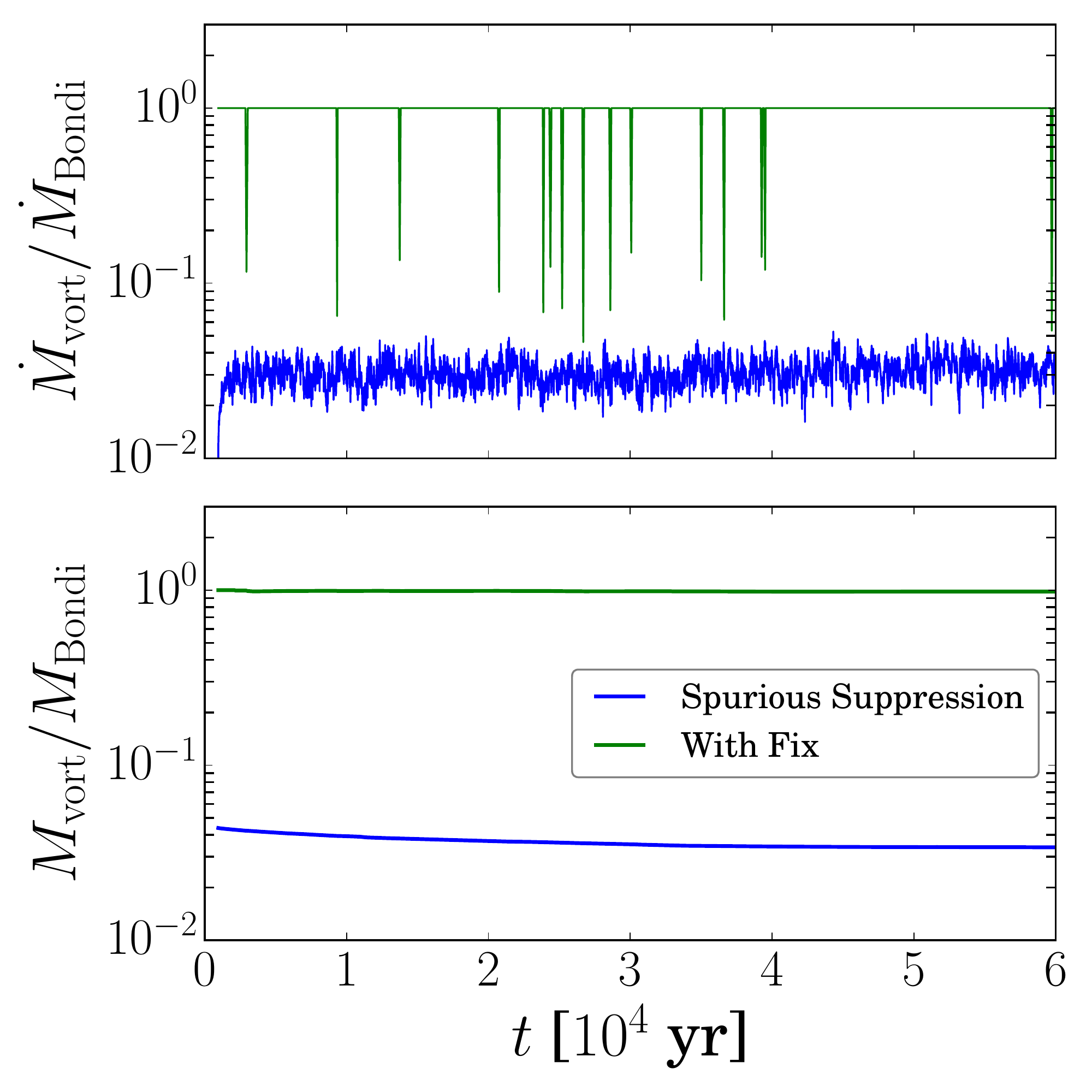}
  \includegraphics[width=0.45\textwidth]{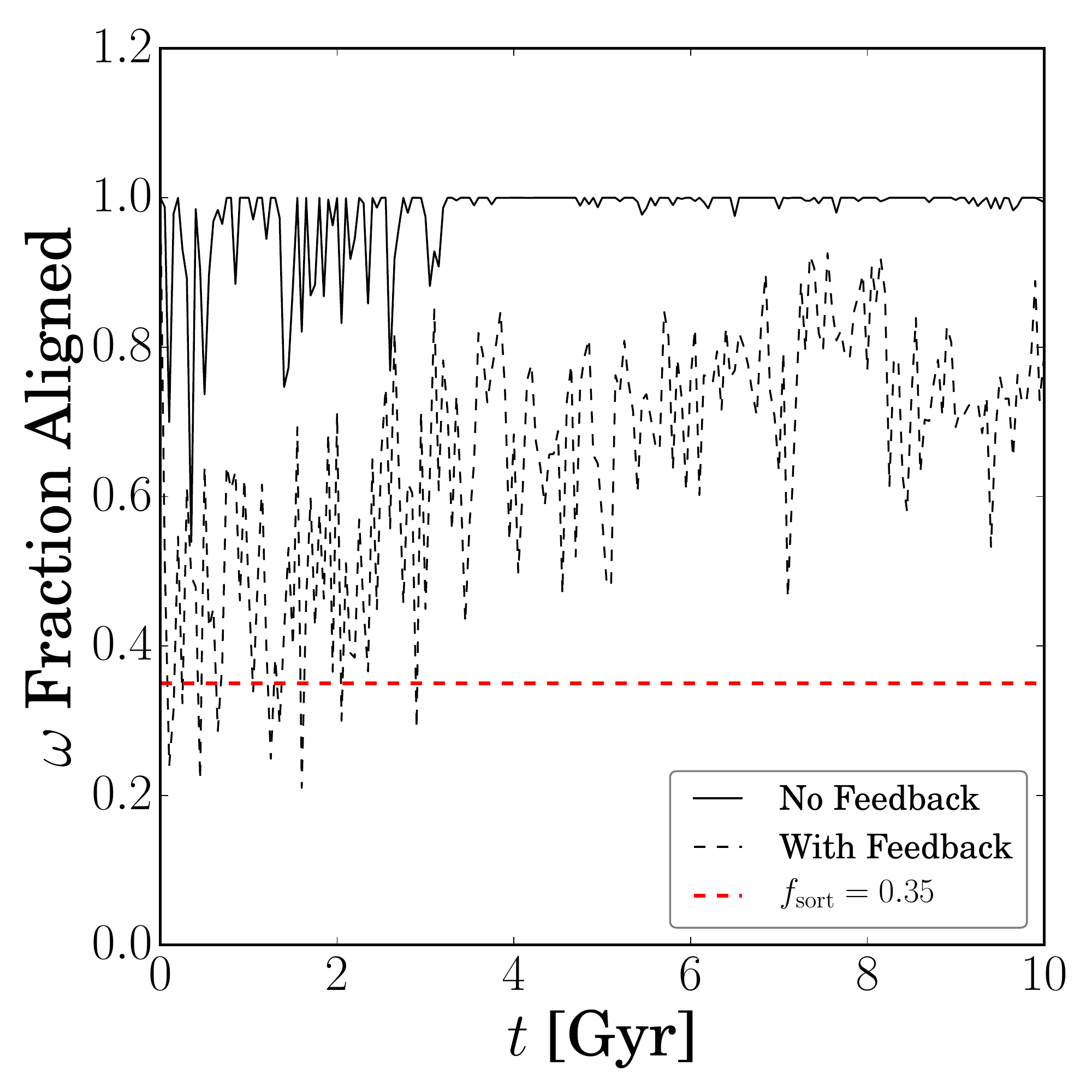}
 \caption{The top row shows the velocity field for simulations of an idealized Bondi
   inflow (left column) and of an isolated disc galaxy with black hole feedback (right column). In the
   second row, we show the corresponding vorticity field for the same
   simulations. In the case of the Bondi simulation, the arrows are normalized
   to a constant value. For context, we show a density slice in the $z=0$
   plane. In the Bondi case, the velocity field is entirely radial, but
   because of the discontinuity of the gradients in the centre of the
   simulation domain coupled with the rising velocity magnitude, the
   simulation results are  increasingly sensitive to small errors in the
   gradient calculation. The resulting vorticity field is entirely noise
   driven, but can lead to a significant spurious signal. In the case of the isolated
   galaxy, the velocity structure of a disc is clearly defined, leading to a
   coherent, robust corresponding signal in the vorticity (aligned with the
   $z$ axis). In the bottom left-hand panels we show both the instantaneous
   and cumulative 
   accretion rate normalized to the analytical Bondi rate with and without our
   $f_\mathrm{sort}$ prescription used to 
   minimize the noisy vorticity estimate. In the Bondi simulation our
   $f_\mathrm{sort}$ prescription is needed and recovers the correct
   accretion rate. In the isolated disc galaxy simulation, both with and without
   black hole feedback, spurious vorticity
   is rarely significant as shown in the bottom right-hand panel, where the
   mass fraction of gas aligned with the net vorticity direction as a
   function of time is plotted.}
 \label{spurious}
\end{figure*}

The \citet{Krumholz} model assumes a large scale disc-like vorticity and, thus
for robustness of our scheme, we need to ensure not to take into account random
turbulent vorticity that may be present in the simulations. This is especially
problematic for radial flows (i.e flows with negligible vorticity) close to
the centre of the simulation domain, since the discrete nature of the velocity
gradient calculation means that the small errors this introduces are
amplified. As this is a function of both the limited resolution of the
simulations, as well as the (Cartesian) coordinate system, we note here that
this problem would not be avoided by using angular momentum (or proxies for
this, such as circular velocity of the gas) instead of vorticity - presence of
minor anisotropies in the radial flow means that, in the central region, it is
inevitable that gas with very large velocities on small but finite impact
parameters will be present.
\label{spurious_vorticity}
In practice, in our simulations with more realistic initial conditions (such
as of galaxies in isolated or cosmological settings), we find
that this should rarely be an issue. Nevertheless, to avoid any spurious
growth suppression, we ensure that the mean vorticity that we measure for the
fluid is representative of a sufficiently large fraction of the gas mass. To
do this, we 
compare the net vorticity within black hole's smoothing length with
that of individual cells within the same region and  
calculate the fraction of the mass $f_\mathrm{sort}$ of the gas that has a
vorticity within $\uppi/4$ of the net direction. For a random distribution of
vorticities, the mean value of $f_\mathrm{sort} \sim 0.15$. By examining the
test simulations, we find that $f_\mathrm{sort} = 0.35$ represents a
conservative threshold which minimizes the vorticity effects that are likely
to be driven purely by noise. As such, for simulations with lower values of
$f_\mathrm{sort}$ we use the standard Bondi accretion rate and not our
vorticity prescription (for more details, see Appendix A). 

Figure~\ref{spurious} shows this phenomenon in practice. We show the velocity
(first row) and vorticity (second row) fields for two simulations, one of a
purely radial Bondi flow (left column) and one with a coherent disc structure
(right column). We have scaled each simulation to the region of interest -
that is, to a region that is comparable to the size of the black hole
smoothing length. In each case, for context, we show a density slice across
the $z=0$ plane. For the Bondi simulation, the vorticity is essentially
completely random, this being just the effect of the noise inherent in
floating point arithmetic. The length of the arrows indicating the vorticity
field have been normalized to the same value: in general, the size of these is
amplified by the speed of the gas, which means that they are negligible except
as the gas approaches the origin, where they blow up in size. Here, because of
the steep velocity gradient, these errors are rapidly magnified, resulting in
a spurious signal that we need to avoid. In the case of the isolated galaxy,
we can see the rotating disc demonstrates a clear vorticity signal, aligned
with the $z$ axis. 

Our simple prescription based on $f_\mathrm{sort}$ works very well in
detecting spurious vorticity and recovering the correct black hole growth
rate, as shown in bottom row of Figure~\ref{spurious}. On the left, we plot
both the ratio of instantaneous simulated black hole accretion rate to the
analytical Bondi rate (top) as well as the ratio of the cumulative simulated
to the cumulative analytical Bondi rate (bottom). Both of these ratios should
be $1$. Vorticity estimation without our $f_\mathrm{sort}$ prescription can
lead to a significant accretion rate suppression, by more than an order of
magnitude, which is purely driven by the numerical noise. Conversely, with our
$f_\mathrm{sort}$ prescription, whilst there are occasions when (by chance)
the vorticity alignment causes the rate to dip for a single time step, this
basically makes no difference to the overall growth, as demonstrated by the
bottom plot. In the right-hand bottom panel of Figure~\ref{spurious} we show
the mass fraction of gas which is aligned to the net vorticity direction
as a function of time for isolated disc galaxy simulation with (dashed) and
without black hole thermal feedback (the results are very similar in the case of the bipolar feedback). This confirms that
spurious vorticity estimate is not an issue in these simulations, as for the
vast majority of simulated time the mass fraction of gas aligned is much higher
than $f_\mathrm{sort} = 0.35$.    

\subsection{Gas Vorticity Measurement in Isolated Disc Galaxy Models}

\begin{figure*}
 \centering
 \includegraphics[width=0.45\textwidth]{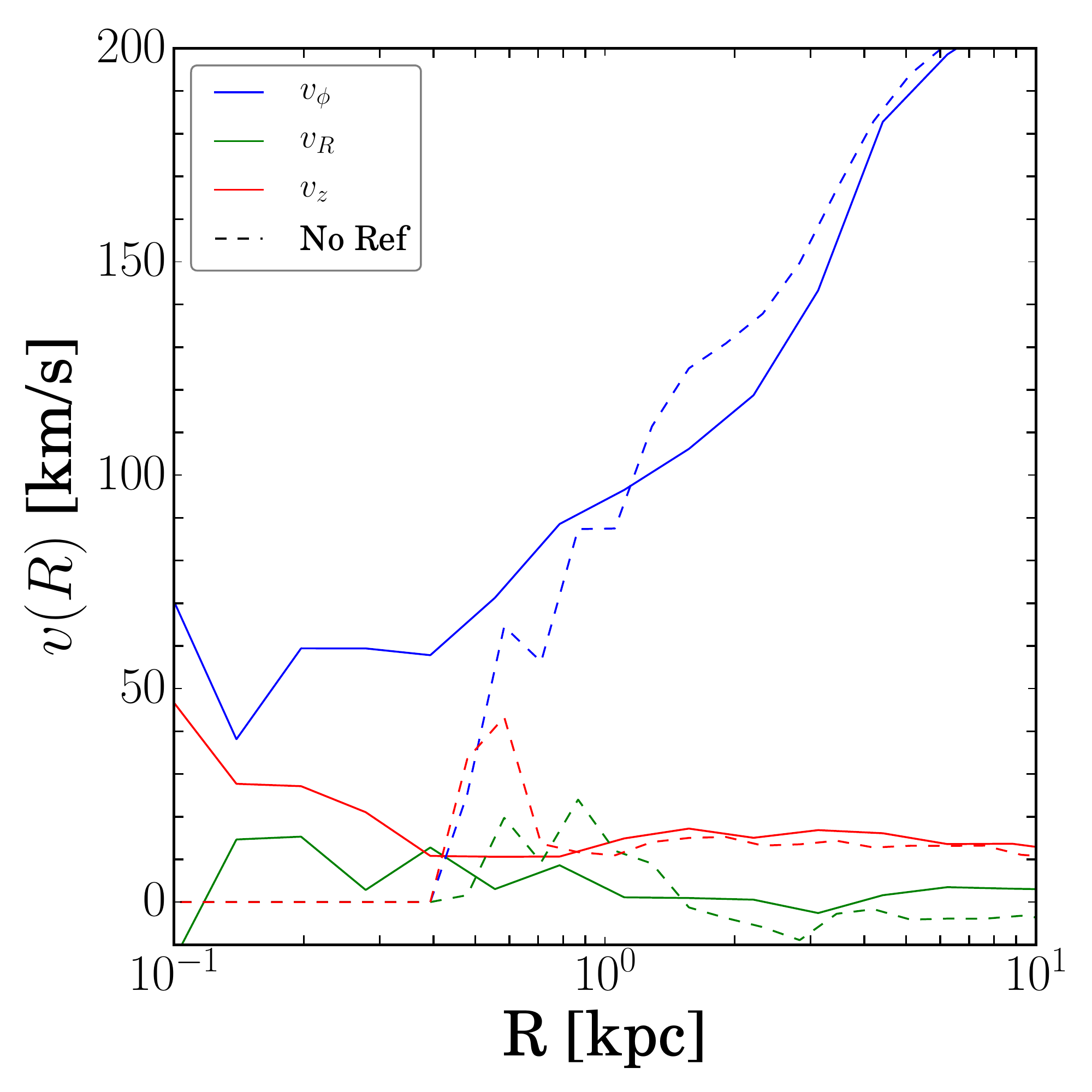}
 \includegraphics[width=0.45\textwidth]{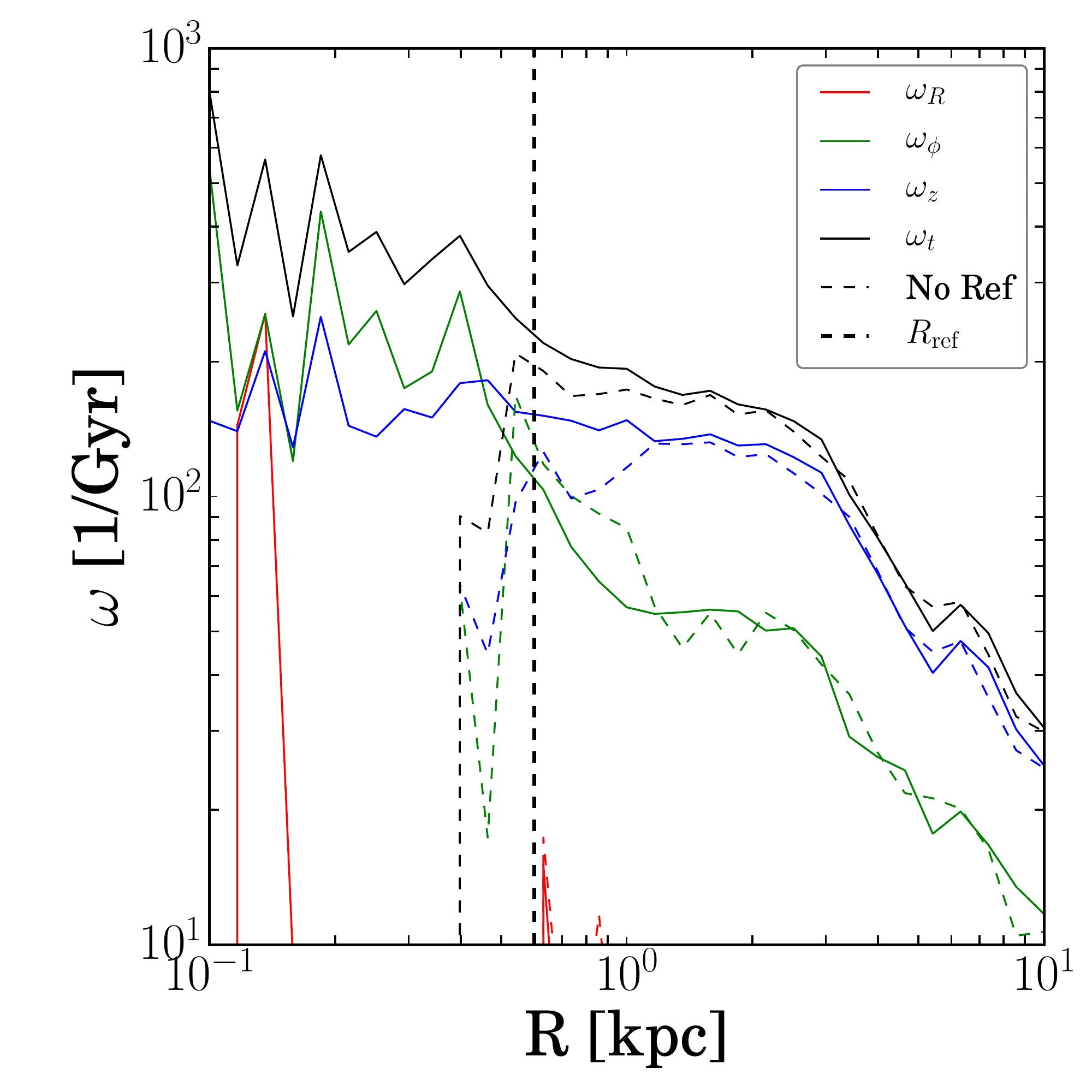}
 \caption{The left-hand panel shows radial profiles of the three velocity components
   in cylindrical (note that $v_z$ represents the absolute velocity value)
   polar coordinates from a simulation with thermal feedback (where the time is
   $1.0\,$Gyr). On large scales, the velocity is broadly similar to the
   initial conditions. On smaller scales, however, there is an increase in the
   vertical velocity, which indicates the presence of an outflow, and the
   radial velocity, as material falls towards the black hole. The right-hand panel
   shows the corresponding vorticity profiles in cylindrical polar
   coordinates. The large scale component is dominated
   by the rotation of the disc (and, as such, the $z$ component of the
   vorticity) whilst on smaller scales, the effects of feedback create a more
   turbulent distribution, with a rising overall vorticity. The vertical dashed
   line indicates the edge of the refinement region. Our simulations
   without the refinement scheme (dashed lines) fail to capture much of this
   complexity.}  
 \label{velocities}
\end{figure*}

\begin{figure*}
 \centering
 \includegraphics[width=0.45\textwidth]{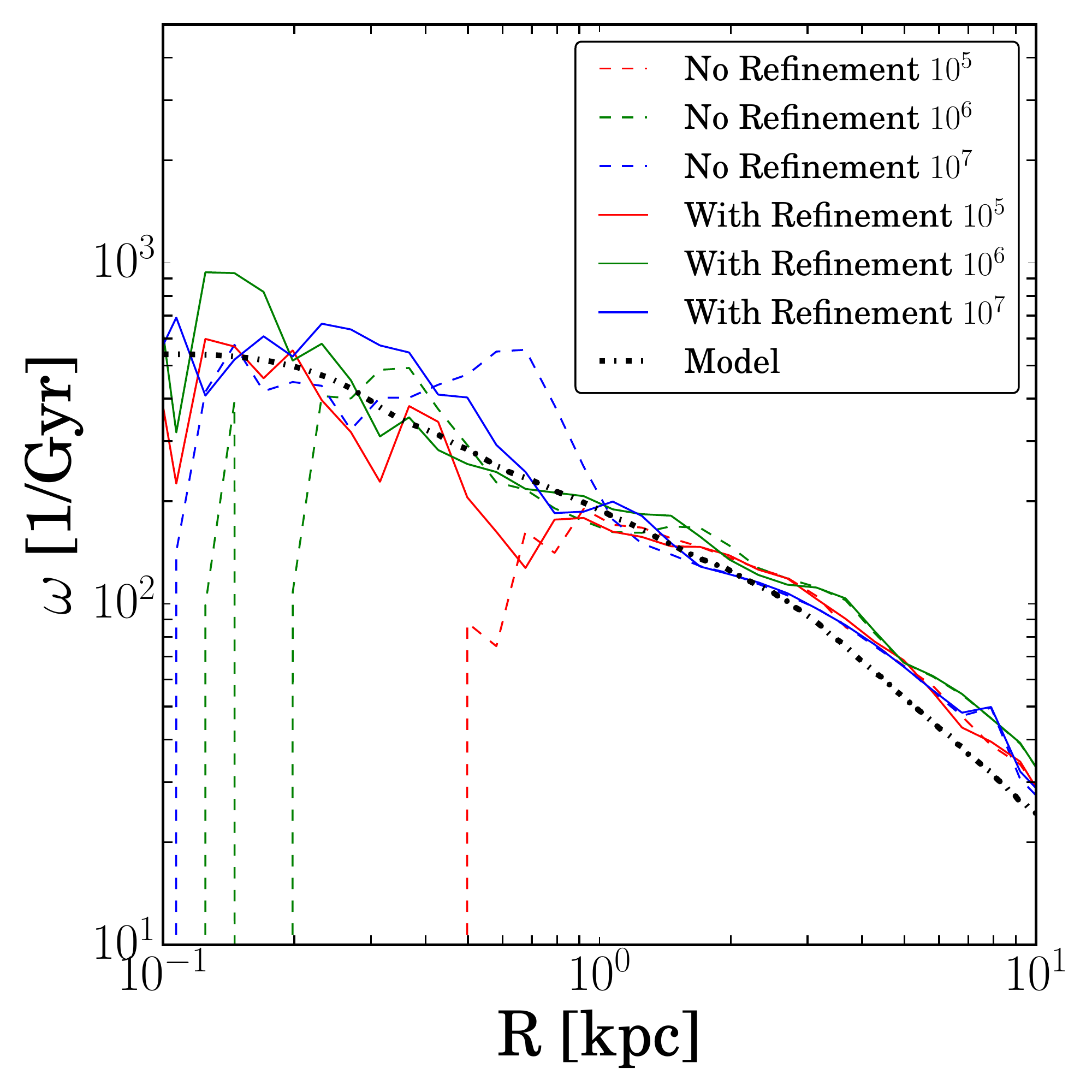}
 \includegraphics[width=0.45\textwidth]{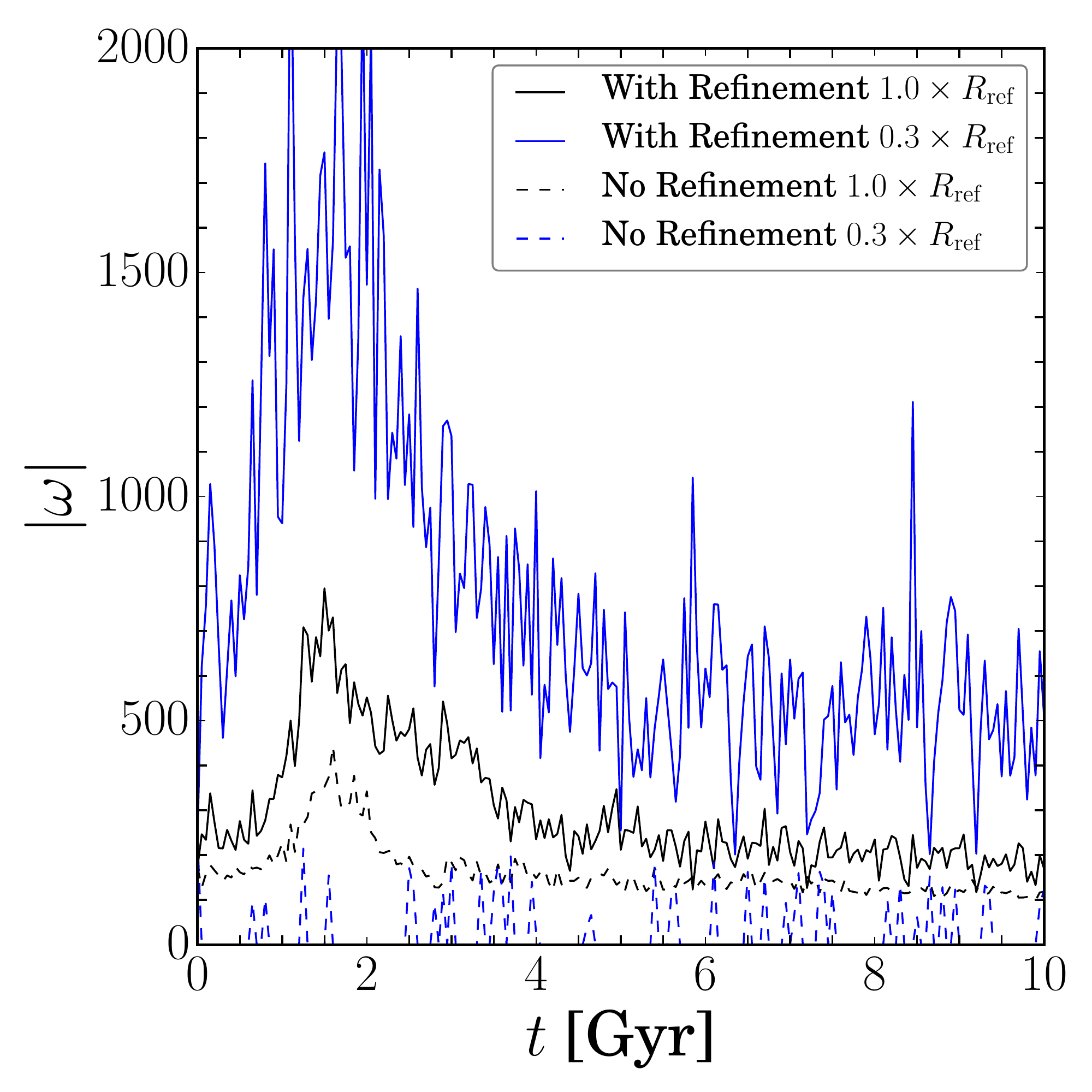}
 \caption{The left-hand panel shows radial profiles of total vorticity for
   simulations with (continuous lines) and without (dashed lines) refinement,
   both performed at three different resolutions. While all simulations have
   comparable vorticity profiles for $R > 1\,{\rm kpc}$, in the inner
   regions only the highest no refinement simulation produces vorticity
   profile which is comparable to all three resolution runs with
   refinement. The thick dot-dashed line shows a simple analytic vorticity profile
   which agrees with the results of our refinement simulations very
   well. The right-hand panel shows the time evolution of the kernel averaged
   vorticity 
 measurement with refinement (continuous lines) and without (dashed). For both
 simulations two examples of estimates are shown: averaging over the full
 refinement region, i.e. a sphere with radius $R_{\rm ref}$ (black lines), and
 over a sphere with radius $0.3\,R_{\rm ref}$ (blue lines). The simulation
 with refinement not only captures a systematically higher vorticity across
 the whole simulated time-span, but it provides numerically meaningful results
 as we probe regions closer to the black hole's Bondi radius, while the
 simulation without refinement fails in this regime at the matching
 effective resolution.}
 \label{vort}
\end{figure*}

All simulations that seek to account for the angular momentum of the gas in
the black hole accretion rate must be able to robustly estimate the velocity
field in the region of interest. Thus to demonstrate the validity of our approach, Figure~\ref{velocities} shows
radial profiles of the three velocity components (left-hand panel)
and the corresponding radial vorticity profiles (right-hand panel) in
simulations with thermal black hole feedback with super-Lagrangian refinement
and without. This shows a problem with simulations without refinement -
because the 
distribution is strongly peaked in the central region, the lack of resolution
here leads to an underprediction of gas vorticity. We note here that this problem
is just a manifestation of the different ways in which the simulations are
capturing the gas properties around the black hole. Previous work
\citep{Curtis:15, Curtis:16} has shown that the increased resolution in the
central region afforded by our refinement scheme can lead to significant
differences in the capturing of the central feedback bubble and the subsequent
transport of the feedback energy away from the immediate vicinity of the black
hole. 

More specifically, feedback has two effects: {\it a)} it causes hot gas bubbles with low angular momentum to rise buoyantly away from the
black hole on vertical (and partly radial) orbits and {\it b)} it increases
the sound speed of the gas, increasing turbulence that will drive viscous
processes that may be responsible in part for transporting angular momentum
outwards and mass inwards. This results in the gas in the central region
having a very different velocity structure. For the cold, accreting gas, cells
fall on to mostly radial orbits, implying $\omega_z \approx {0}$, resulting in
the drop in vorticity for the smallest radii. In addition, however, the strong
black hole-driven outflow means that $v_z$ has an $R$ dependence in the
innermost region. This results in a strong contribution to the vorticity from
the azimuthal $\omega_\phi$ term.

The left-hand panel of Figure~\ref{vort} shows the radial profiles of total gas
vorticity for simulations with and without refinement at three different
resolutions. The thick dot-dashed line is a simple analytic estimate of the
vorticity profile which we calculate as follows: we assume that the dominant
component in the vorticity signal is aligned with the $z$ axis of the disc and
is dependent on the circular velocity, which we approximate as $\sqrt{GM(<R) /
  R}$. We then differentiate this field numerically to find the curl and,
hence, the vorticity. All three refinement simulations at increasing
resolutions produce 
very similar vorticity distribution and agree very well with our analytic
estimate. Simulations without refinement, however, fail to reproduce inner
vorticity peak a part from the highest resolution run. This indicates that for
a comparable effective (large-scale) resolution and CPU resources, simulations
without refinement are bound to systematically underestimate gas
vorticity. This is illustrated explicitly in the right-hand panel of
Figure~\ref{vort} where kernel averaged vorticity measurement is plotted as a
function of simulated time. Not only is the estimated vorticity systematically
smaller in the runs without refinement, but as we probe regions closer to the
Bondi radius the estimate is dominated by the numerical noise (dashed blue
line), demonstrating that much higher effective (and computationally more
expensive) simulations would be needed to mitigate this issue.  

\section{Simulation results: Implications for black hole growth} \label{Implications}
\begin{figure*}
\centering
\includegraphics[width=0.45\textwidth]{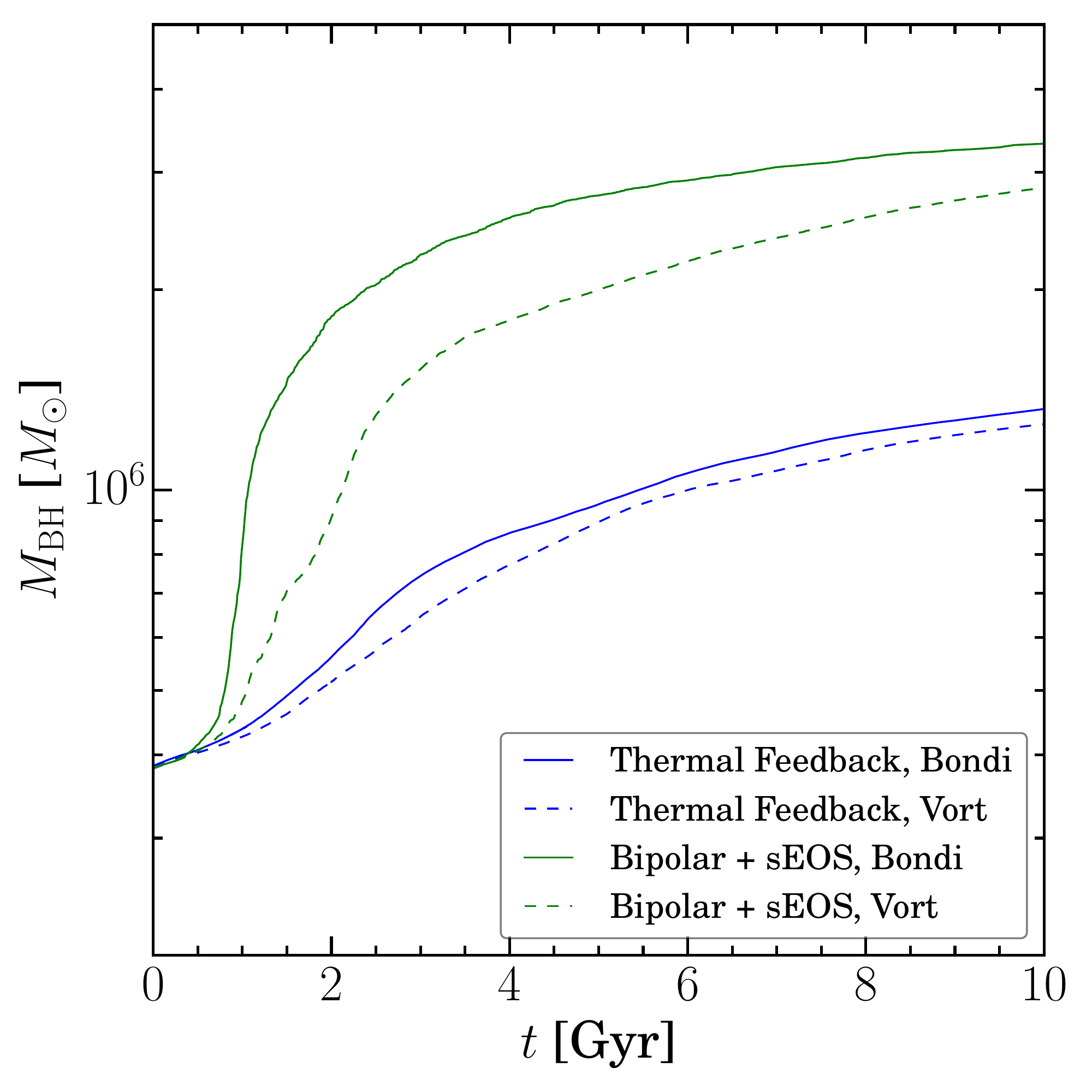}
\includegraphics[width=0.45\textwidth]{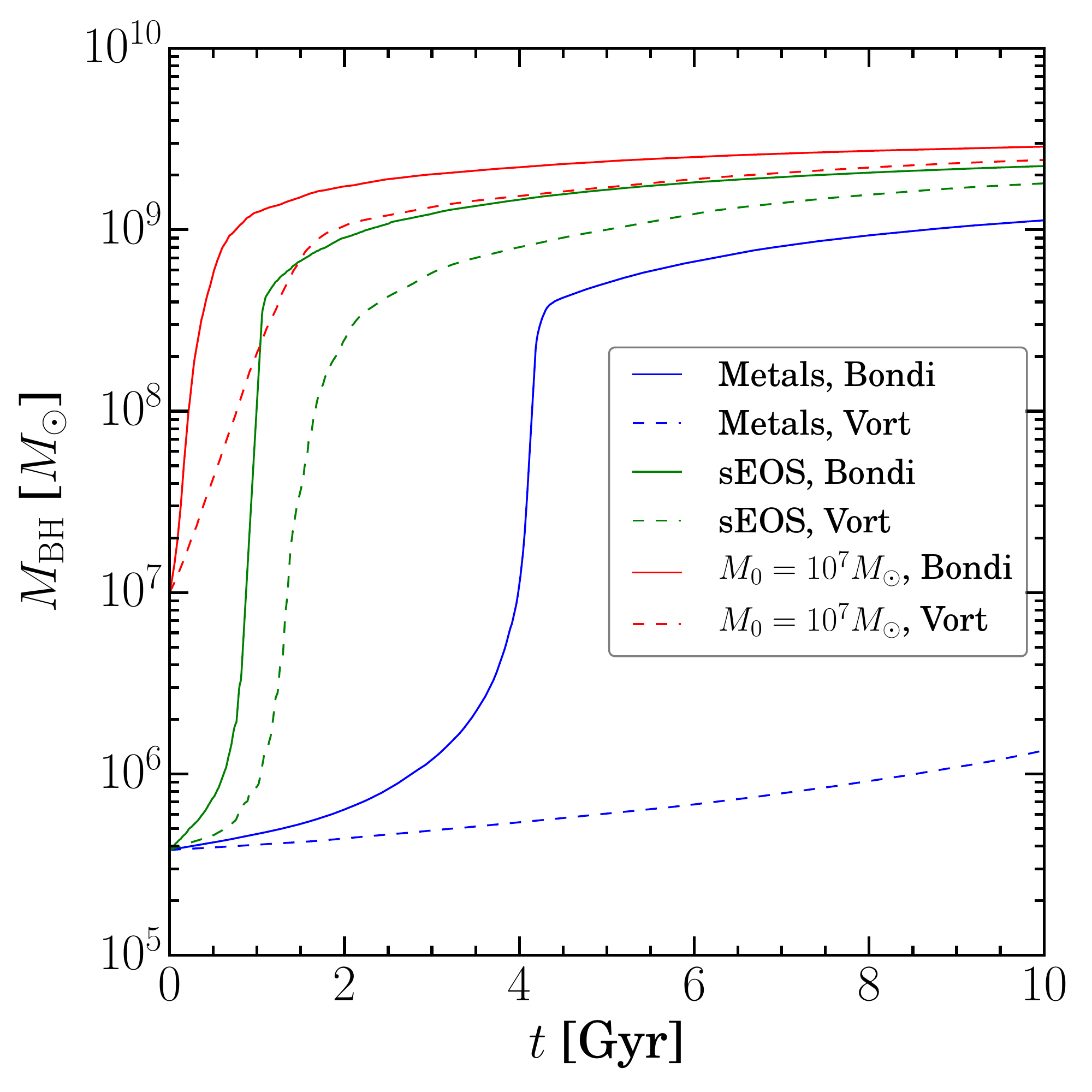}
\caption{The left-hand panel shows the growth of black holes in simulations with
  two feedback prescriptions: thermal and bipolar coupled with a sEOS, taking
  into account the vorticity-based and standard Bondi accretion prescriptions.  In
  the case of the thermal feedback, the sound speed is too high to allow for
  significant suppression by vorticity. In the bipolar case, whilst the black
  holes reach a similar final mass (which is determined in a large part by the
  cumulative energy injected by the feedback) there is a delay in the initial
  growth of the black hole of up to $\sim 1$Gyr. The right-hand panel shows the
  growth of black holes in simulations without feedback. In these cases,
  despite the black hole growth frequently being in the regime of significant
  suppression due to vorticity, the accretion rate is sufficiently high that
  the black holes grow quickly regardless. The exception to this is the simulation
  with metal line cooling, where all the relevant parameters remain in
  the critical region, as discussed in Section~\ref{analytic_distributions}
  (see Figure~\ref{delay}).} 
\label{mbh} 
\end{figure*}

\subsection{Isolated disc galaxies}

Figure~\ref{mbh} shows the effects of our vorticity prescription on the
evolution of the black hole mass in the isolated disc galaxy simulations. In
the left-hand panel we show the results for simulations with black hole
feedback. Here, 
we can see 
that for standard isotropic thermal feedback, as expected from our analytic
analysis, the suppression of the black hole growth is minimal. The surrounding
gas is at too high a sound speed and as highlighted by Figure~\ref{suppression}
this simulation model probably represents the worst case for vorticity-based
accretion suppression to be effective.

For comparison, we also show a simulation run using bipolar feedback, which has
the effect of decoupling (to a certain extent) the black hole heated gas from
the cold accreting gas. Here, we adopt a softer equation of state for the
ISM as well, leading to a less pressurized and more clumpy medium. The
relative 
suppression of the black hole growth is more significant  - indeed, at its
peak $\dot{M}_{\rm Vort}$ is around one tenth of the standard Bondi rate - and
has a noticeable effect especially at early times, with the peak growth phase
of the black hole delayed by around 1 Gyr in the vorticity-based simulation. Note however that the final black hole masses are broadly independent of the
accretion rate prescription (and much more dependent on the feedback
prescription) as black holes enter into self-regulated growth regime in both
of these simulation models. Black hole growth is thus essentially determined
by the amount of injected energy. In these cases, therefore, the suppressing
effect of gas vorticity is sub-dominant compared to the feedback choice.

Given the strong dependence on feedback, we now turn to investigate the
effect of the suppression in simulations with no feedback, in order to
restrict our analysis to the effects of the accretion rate alone. 
Whilst such simulations have been shown to be unrealistic in many ways, they
nevertheless provide insight into the different regimes of black hole growth
in a simplified scenario. Specifically, while it is desirable on both
theoretical and numerical grounds for the black hole growth to be
self-regulated, in reality this is not necessarily the case, or it may occur on
average with short, intermittent phases of rapid, unrestricted growth in
between. Thus, our simulations without feedback and with strong
self-regulation may each possibly book-end a more realistic scenario. 

In the right-hand panel of Figure~\ref{mbh}, we show the evolution of the
black hole mass for simulations without feedback, for three different setups:
a fiducial simulation (which includes metal line cooling), a simulation with a
softer equation of state and finally a simulation with a larger initial
black hole mass of $10^7 M_\odot$. Here, we are motivated by exploring
different regimes that may lead to significant suppression, according to our
previous analysis in Section~\ref{analytic_distributions}. Black holes in
simulations with a higher initial seed mass and with the softer equation of
state grow rapidly, showing a very modest difference between the standard and
vorticity simulations. This is despite the fact that in both cases, at its
peak, the ratio of the vorticity-based prescription to the standard Bondi rate
is around $0.01$. The reason for this is that, in both cases, the fast black
hole growth is driven by changes in the Bondi-part of the accretion rate (also
due to a strong $M_{\rm BH}^2$ dependence) leading quickly to the
Eddington-limited growth, to  
the extent that even a significant vorticity-based suppression has a limited
effect on the overall evolution. For the simulation with metal line
cooling, however, it is possible to enter a regime where the black
hole growth is significantly suppressed by roughly three orders of magnitude and
for the entire 10 Gyr of simulated time-span. While this highlights that
gas angular momentum barrier can be extremely efficient in limiting black hole
growth, we stress that the relevance of this result is clearly
highly dependent on the actual gas properties on parsec scales and below,
which are currently not well understood.

\subsection{Merging Galaxy Pairs}
\begin{figure*}
\centering
\includegraphics[width=0.24\textwidth]{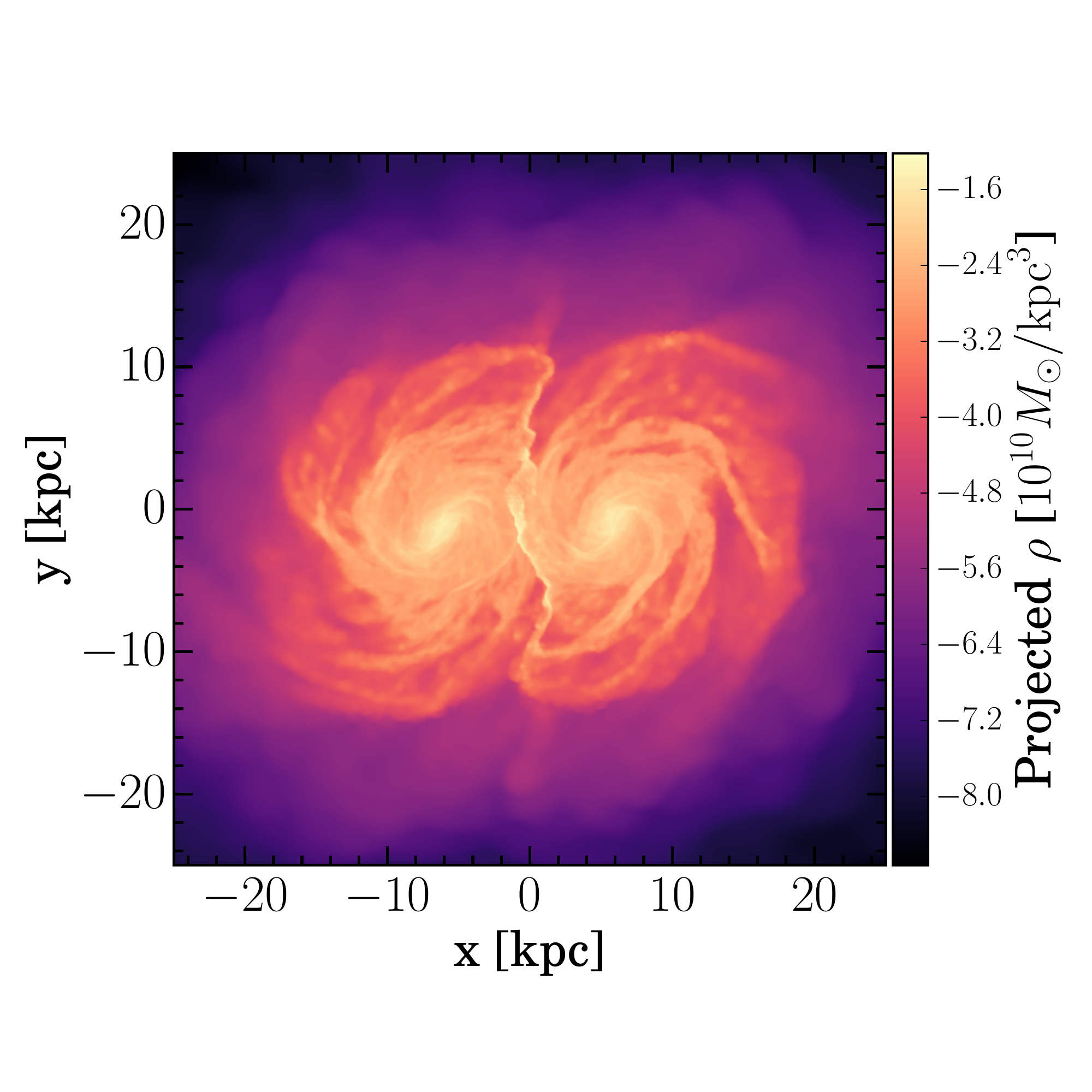}
\includegraphics[width=0.24\textwidth]{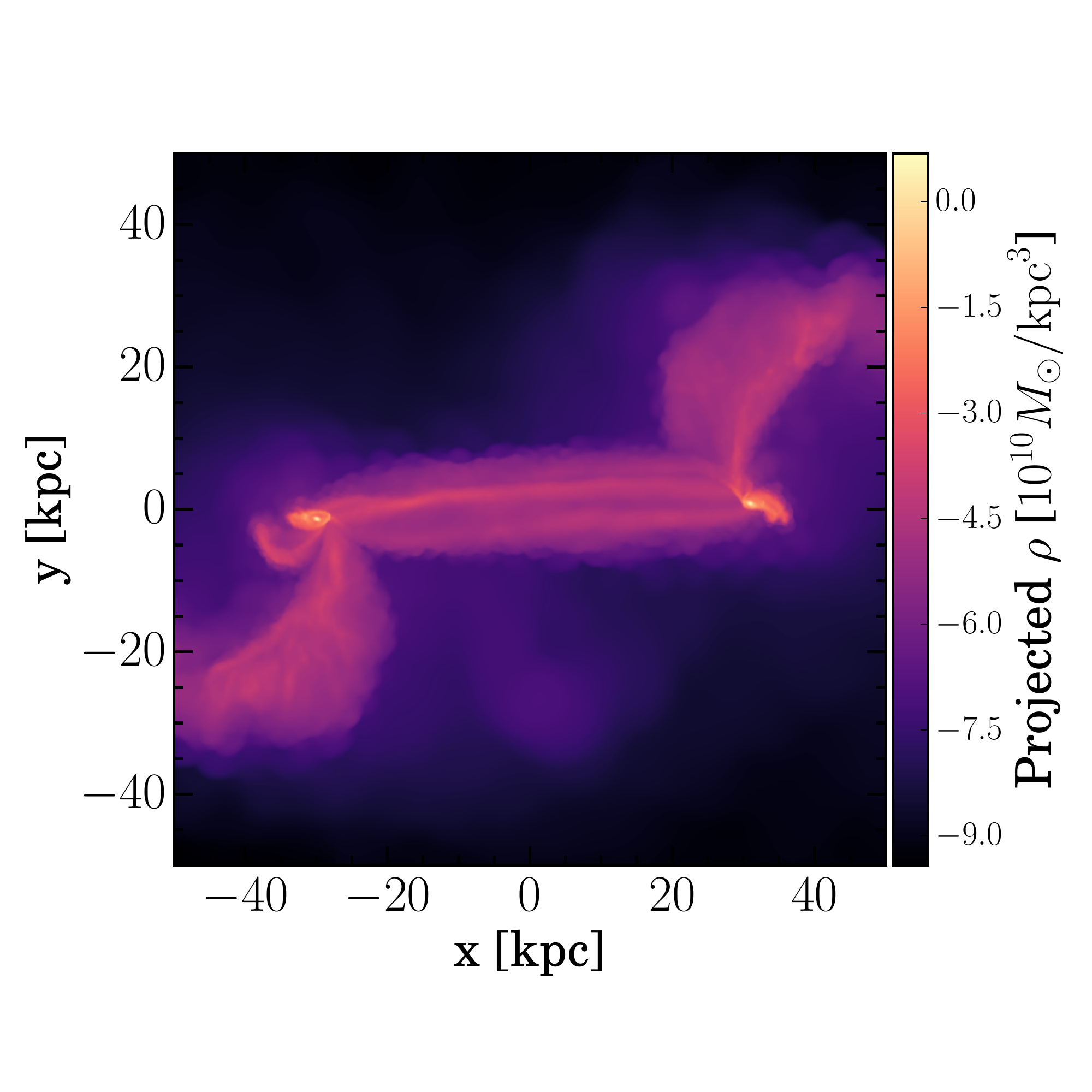}
\includegraphics[width=0.24\textwidth]{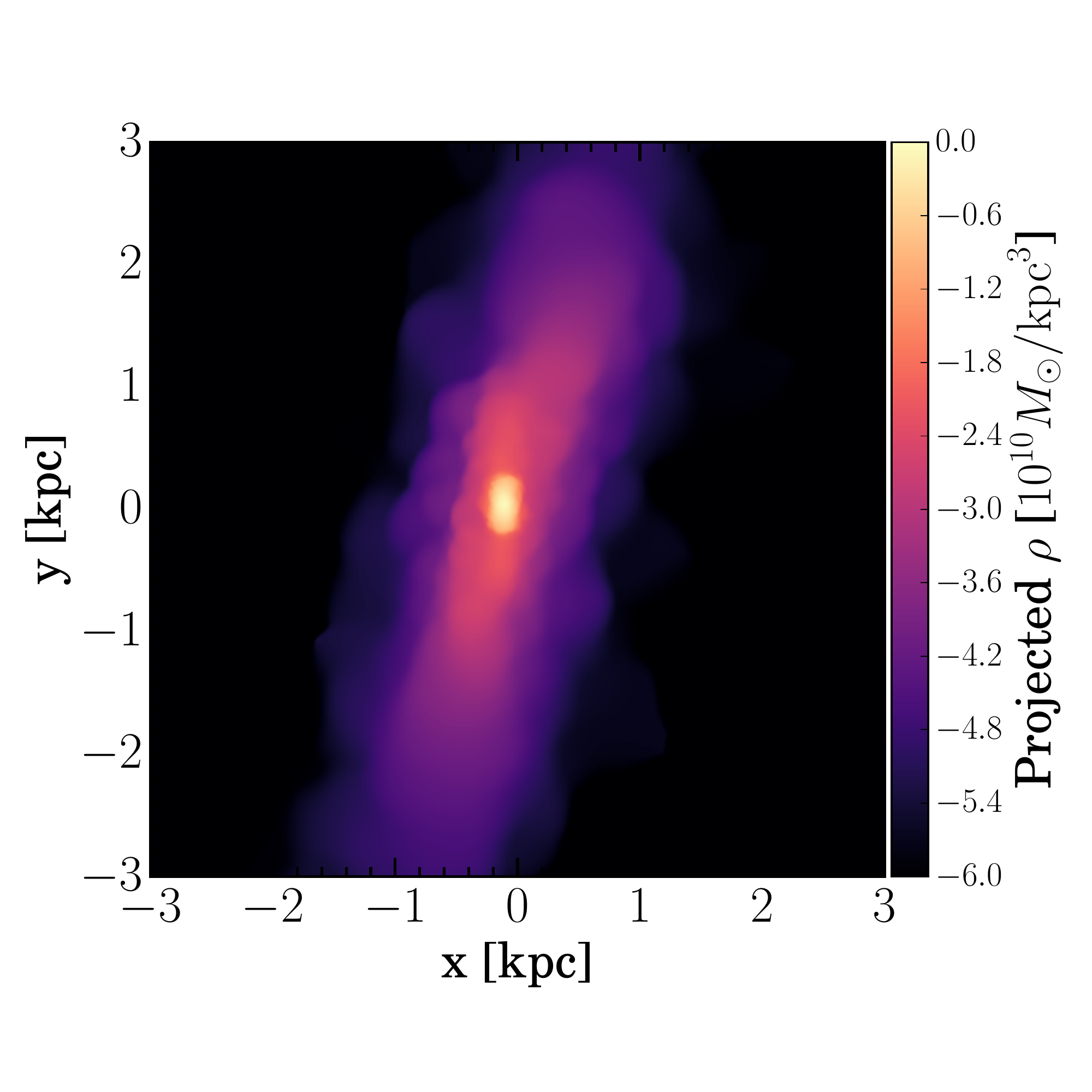}
\includegraphics[width=0.24\textwidth]{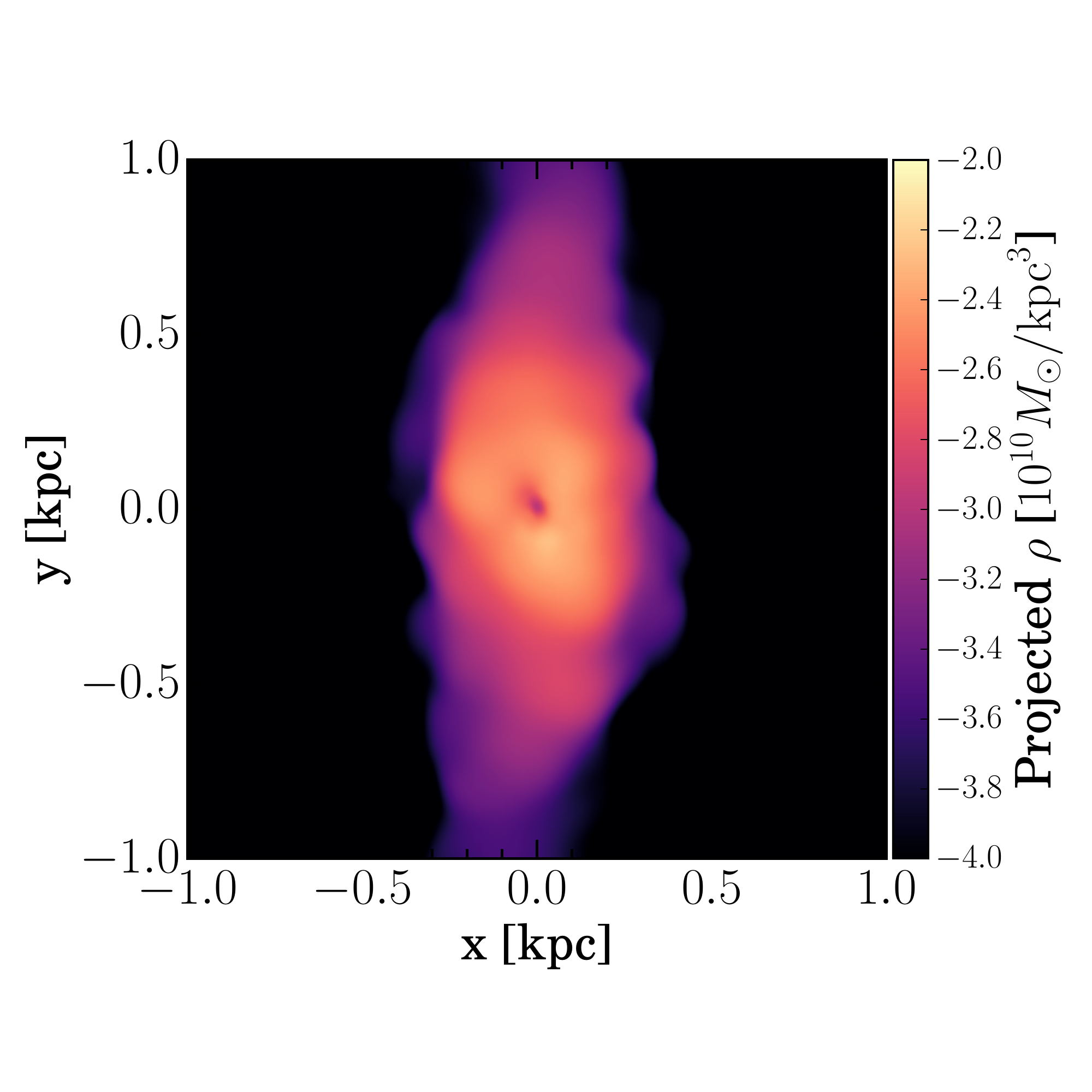} \\
\includegraphics[width=0.24\textwidth]{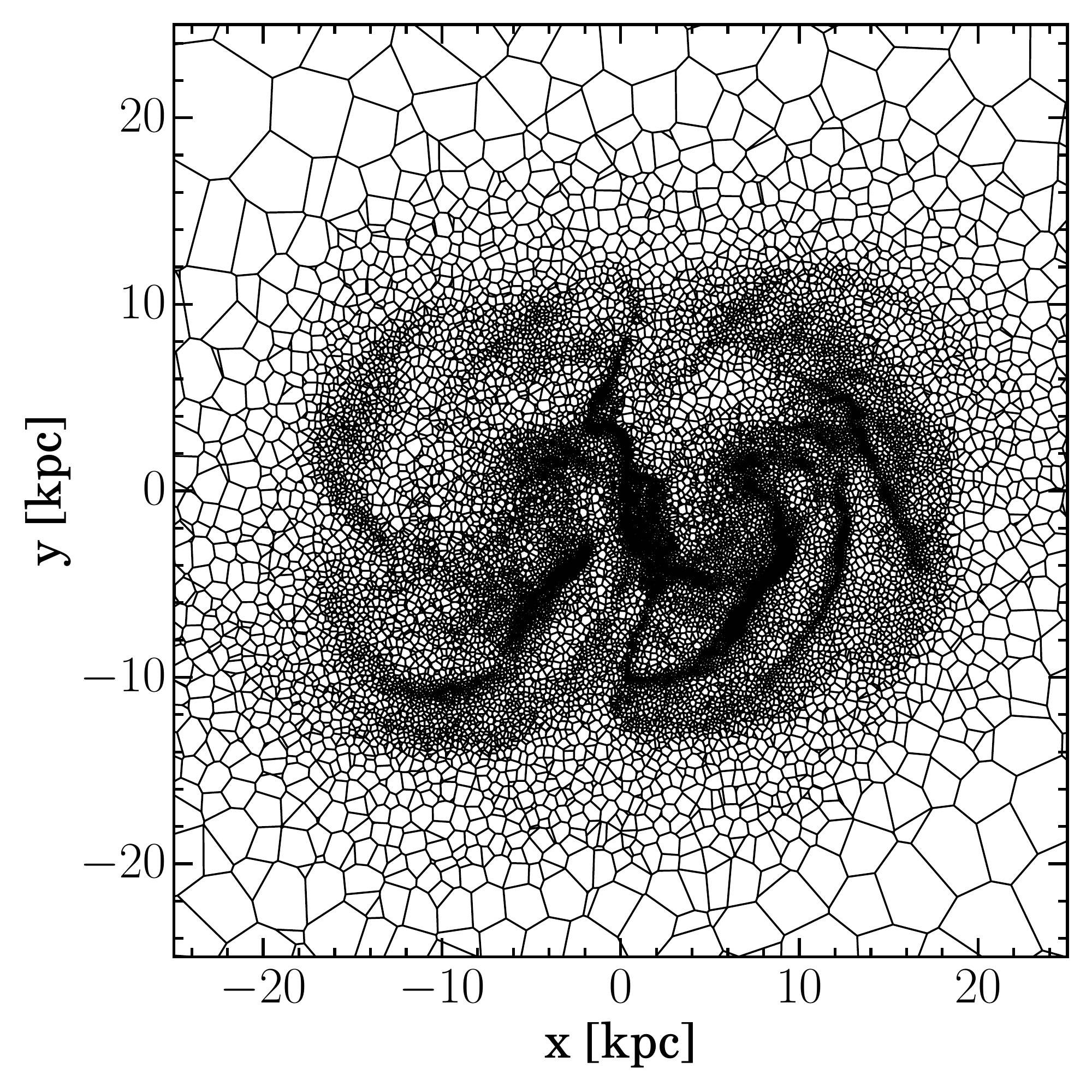}
\includegraphics[width=0.24\textwidth]{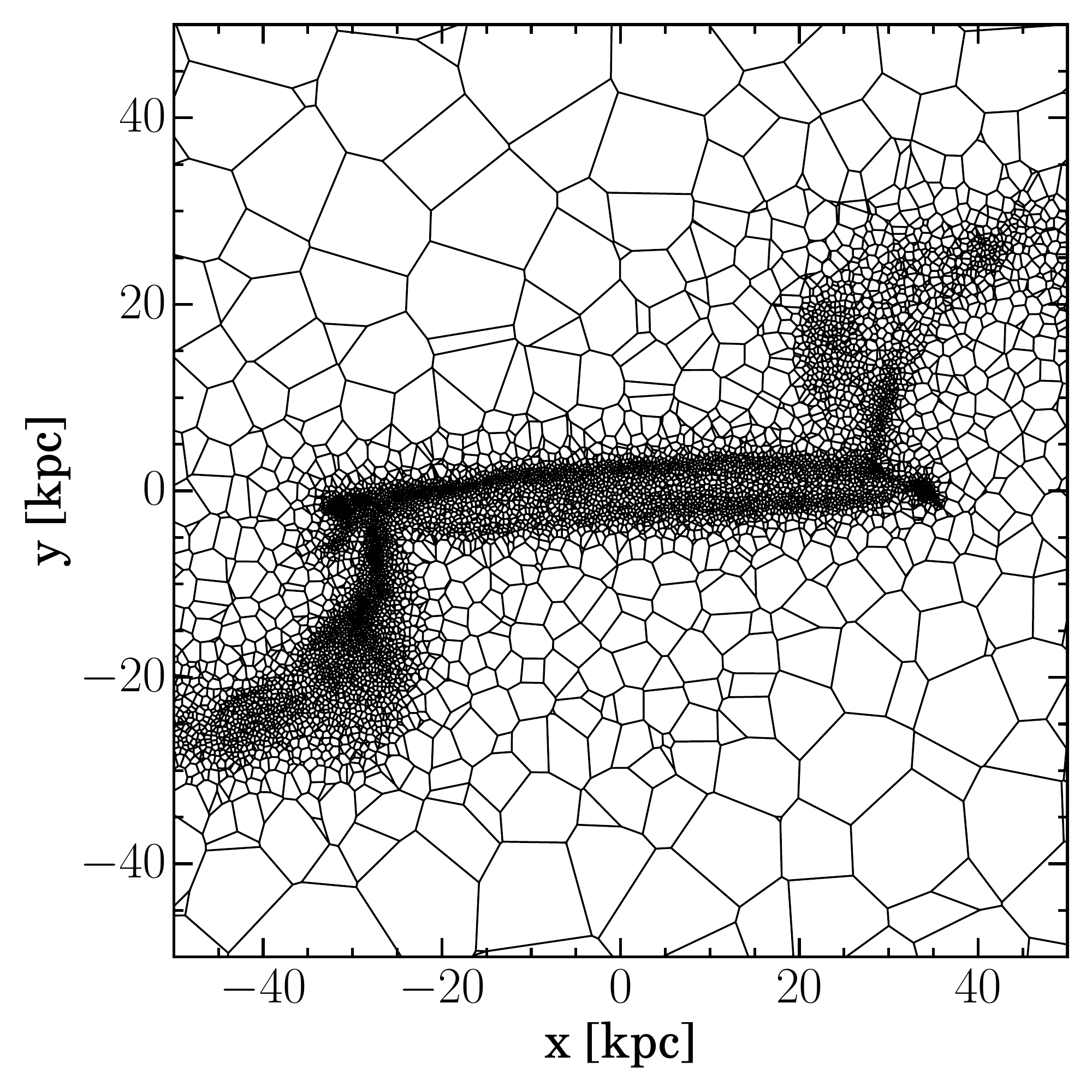}
\includegraphics[width=0.24\textwidth]{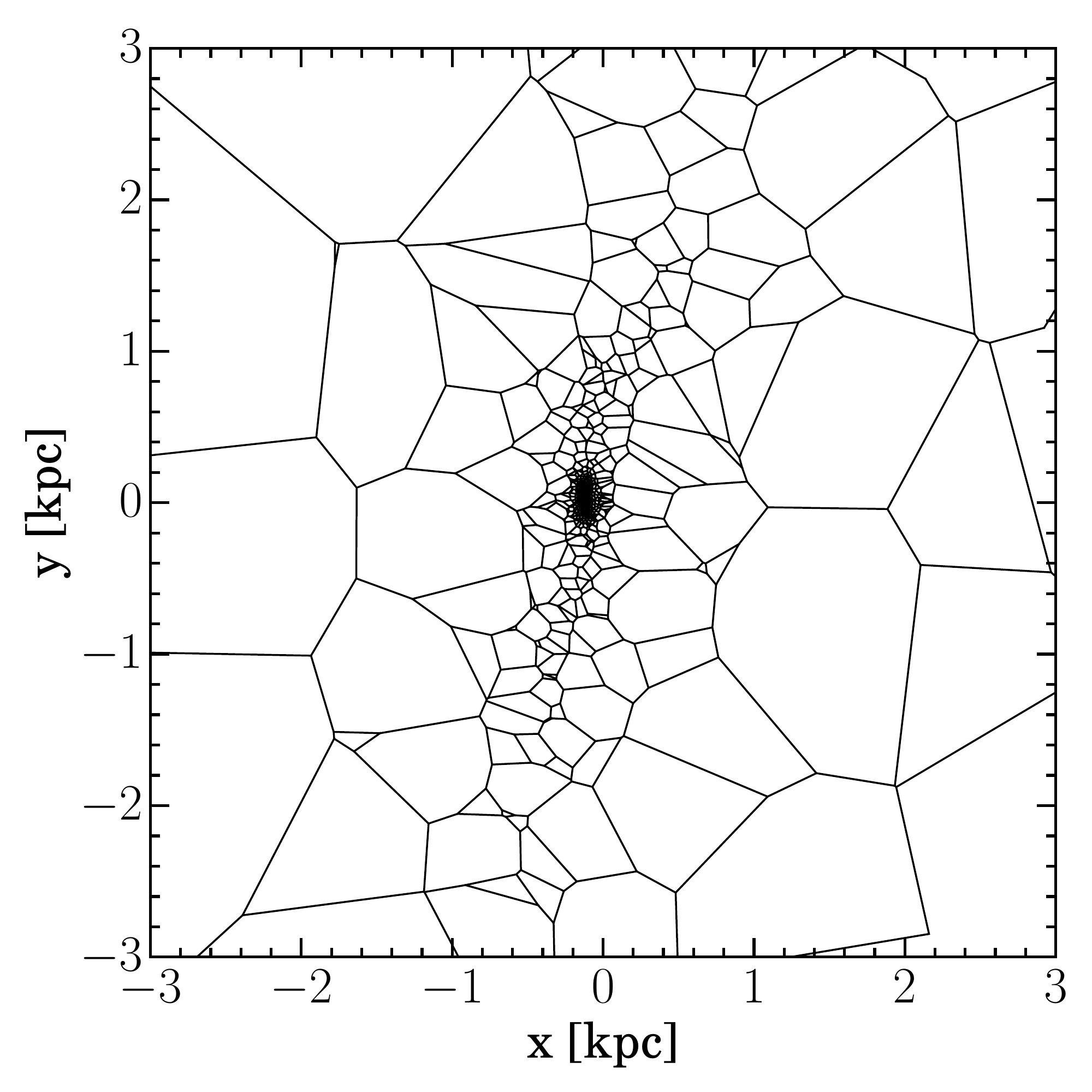}
\includegraphics[width=0.24\textwidth]{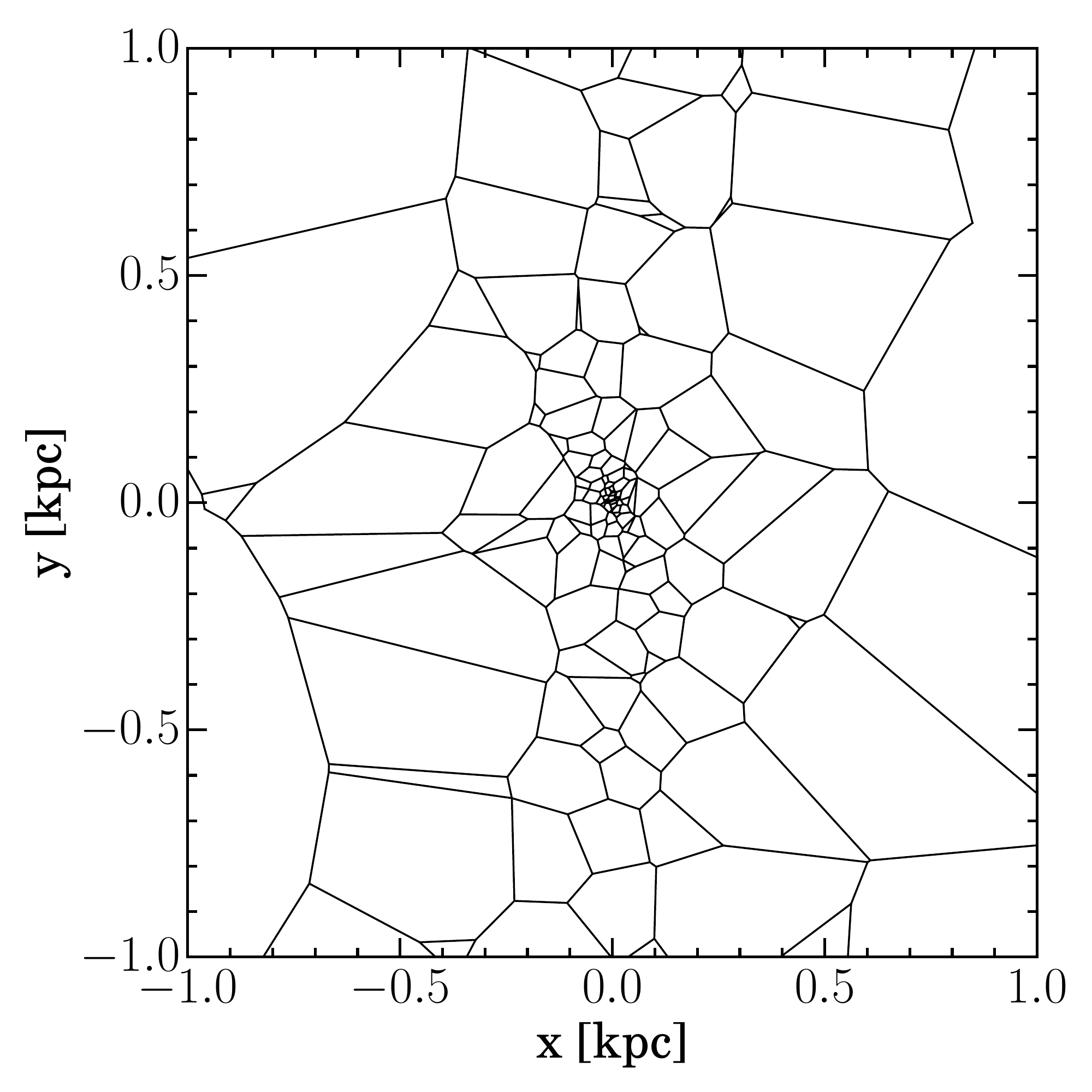} \\
\includegraphics[width=0.95\textwidth]{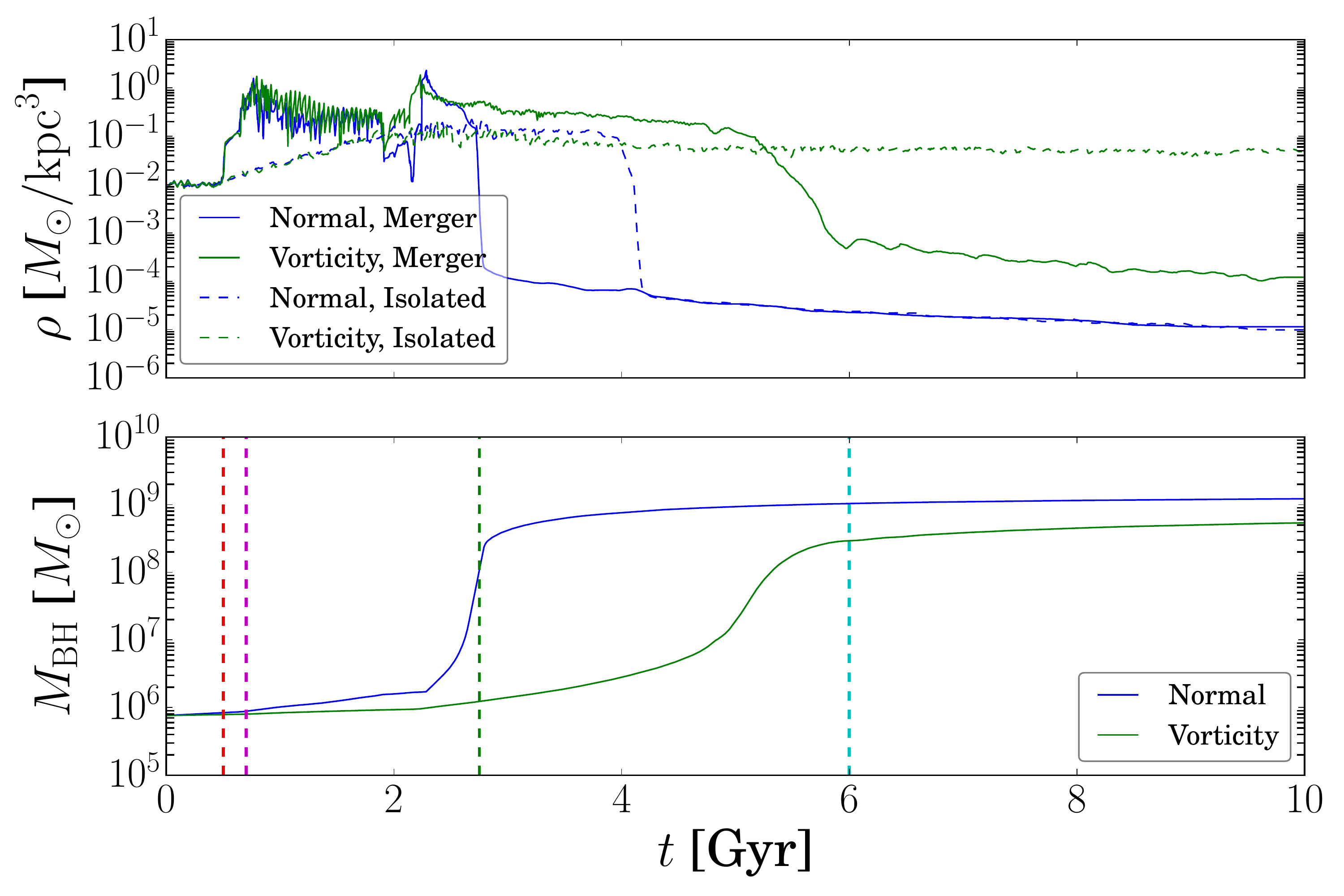}
\caption{The top row shows the projected
  gas density of our simulations of a galaxy merger pair at times of $0.5$, $0.7$, $2.75$ and $6\,$Gyr (from left to right), spanning a
  time sequence from the first contact to the final merger remnant. The second row
  shows the corresponding Voronoi mesh, which illustrates the spatial
  resolution of our hydro solver. Below this, the evolution of the gas density
  within the black hole's smoothing length is plotted, including the
  case with the standard Bondi accretion prescription (blue) and with the
  vorticity-based prescription (green). For comparison we plot the same
  quantity from our simulations of isolated disc galaxies simulated with the
  same physics (dashed lines). Finally, the bottom row shows the evolution of the
  black hole mass for our merger simulations with Bondi and vorticity-based
  prescriptions (dashed vertical lines correspond to the image sequence shown in the top row). While there is a significant delay in black hole growth after
the second passage with vorticity-based prescription, due to the large-scale
torques high gas central density builds up, which ultimately leads to the final
black hole mass similar as in the standard Bondi case.}
\label{merger} 
\end{figure*}

In order to study the effects of our vorticity-based prescription on a
situation where the angular momentum profile of the gas in the galaxy is
significantly disrupted, we perform simulations of galaxy binary
mergers. Here, we set up two galaxies with the same parameters as those in
Section~\ref{ICs} at a distance of $200\,{\rm  kpc}$, which places them
outside their respective virial radii. The two galaxies are initially in the
same plane and collide on a prograde parabolic orbit. We limit ourselves here
to studying the most interesting setup from our original simulations that
gives a significant suppressing effect, namely one with no black hole feedback
and with metal line cooling. 

As the merging galaxies approach each other, they slow down due to
dynamical friction and after two close passages they coalesce to form a
spheroidal remnant\footnote{Note that due to the relative orientation of galaxies in this specific merger setup, a gaseous disc reforms during the intermediate stage of the merger, which has considerable vorticity.}. Large scale gravitational torques during the merging
process funnel gas towards the innermost regions thus increasing the gas
density around black holes as well as increasing the relative fraction of gas
on radial versus azimuthal orbits. In the top row of Figure~\ref{merger} we
show a selection of projected gas density maps from different stages of the
merger evolution. The corresponding Voronoi tessellations of the computational
domain are plotted in the second row. The first galaxy passage happens at
around $0.5\,$Gyr, followed by a second passage at around $1.5\,$Gyr, leading
to the binary black hole coalescence at around $2\,$Gyr.

During the merger a significant amount of gas is driven on to low angular
momentum orbits, as illustrated in the third row of Figure~\ref{merger}. Here,
we plot the gas density measured within the black hole's smoothing length as a
function of time. For comparison, we also show gas density from the isolated
disc galaxy simulations using the same physics. The initial gas density
evolution is identical, but at the first merger passage a large influx of gas
increases the central density by around two orders of magnitude in the merger
runs, which drives initial black hole growth in the standard Bondi
simulation. Note that in the corresponding simulation with vorticity-based
accretion rate prescription this initial black hole growth is suppressed, as
gas has still a moderate vorticity. Large-scale
torques during the second passage drive yet more gas towards the central
region not long after the final merger of the two black holes, and this leads
to a period of rapid, Eddington-sustained growth in the simulation with the
standard Bondi rate. As black hole growth is not feedback regulated, central
gas density quickly drops in the Eddington phase as the black hole is
efficiently swallowing its surrounding gas reservoir. 

This considerable angular momentum transport also has the effect of allowing
the black hole with the vorticity-based accretion rate prescription to grow to a
significantly higher mass than in the isolated disc galaxy case, as the
central gas density stays about an order of magnitude higher. However, the black
hole growth is modulated by the presence of non-zero gas vorticity. This
leads to a delay of more than $2\,$Gyr before the Eddington-limited phase is
reached, which ultimately leads to the final black hole mass similar to, albeit
somewhat smaller than, the standard Bondi case. This effect is very
interesting as it leads to a large time-offset between the peak in
star formation activity (which occurs at $0.7\,$Gyr) and the peak in black
hole growth and thus the occurrence of the quasar phase. Taken at the face value, this could
also explain the observational lack of evidence that the quasar triggering
occurs in host galaxies with perturbed morphologies: in the vorticity-based
prescription major black hole growth occurs at around $5.5\,$Gyr when the
galaxy merger remnant is largely relaxed and spherical (see right-hand top panel
of Figure~\ref{merger}).  

Whilst during the merger large-scale torques clearly channel much of the gas
on to low angular momentum orbits, allowing it to move to within the accreting
region of the black hole, we find that the effects of this are not significant
enough to completely overcome the suppression encoded in our vorticity-based
prescription, i.e. that the inner gas flow is not purely radial. This is,
perhaps, unsurprising since the condition in \ref{very_small_omega} is
especially stringent. Although to a certain extent, we are limited by our finite resolution and
numerical precision given the relatively tiny size of the Schwarzschild radius,
more generally it is worth stressing that the validity of our results is
restricted by poorly known gas properties on spatial scales of parsecs and
below, where gas fragmentation, star formation and their associated feedback, as
well as the nature of the gas flow on even smaller scales (e.g. in the
``alpha-disc'' regime) may dominate the final supply rate to the black hole.  

\section{Conclusions} \label{Conclusions}

In this paper we have investigated the impact of gas angular momentum on the
growth of supermassive black holes. Specifically, we have limited our analysis
to the model initially proposed by \citet{Krumholz} that takes into account
gas vorticity as to generalize the well-known Bondi accretion rate. While by
no means exhaustive, this model is well suited for implementation in galaxy
formation simulations, especially in the case where the relevant gas
properties are robustly measured close to the Bondi radius, which is the case
when we adopt our super-Lagrangian refinement technique \citep{Curtis:15}. Note
that the refinement is not only needed to accurately measure gas density and
sound speed, but the gas vorticity field itself, which is even more
numerically challenging. We have implemented this vorticity-based accretion
rate prescription in the moving mesh code {\small AREPO}, for general usage in
(cosmological) galaxy formation simulations. We have extensively tested our
implementation on a range of numerically challenging setups obtaining robust
results and have verified the convergence properties of the model with
increasing resolution.  

With respect to the Bondi rate, in the vorticity-based prescription a
significant suppression of black hole growth is expected when gas angular
momentum and thus vorticity are high. To fully investigate model
predictions we have first developed simple analytical dynamical models,
which allowed us to establish a tight dependence of the
relative suppression magnitude on the density and sound speed of the gas - two
properties that on the scales of interest are poorly constrained in practice
by both observations and 
simulations. Moreover, due to the significant dependence of the accretion rate
on the black hole mass ($\propto M_{\rm BH}^2$), our dynamical models reveal a
more nuanced effect, whereby significant accretion suppression can be
outweighed by rapid growth in cases where the net accretion rate is high
(i.e. close to the Eddington limit).
  
We have then explored the effects that different black hole feedback
parametrizations and ISM physics choices have on the black hole growth with
the matching set of simulations of isolated disc galaxies, performed with the
standard Bondi and vorticity-based accretion rate prescriptions. In doing so,
we have 
found a picture that is broadly similar to our analytic predictions: black
hole growth is largely governed by feedback but that, within this, the effect
of the vorticity prescription is to have a mild suppression on the accretion
rate. One of the principal reasons for this is the self-regulating nature of
feedback implementations we considered in this study, where the final black
hole mass is largely set by the amount of feedback energy injected.

Conversely, in simulations without black hole feedback, we found that the
black hole can be in regimes where the suppressing effect is much more
significant, even leading to no appreciable growth over a Hubble time. The
contrasting no feedback and feedback simulation results are likely to bracket
a more realistic scenario, where black hole self-regulation is not as tight
as we assumed here and occurs on a natural duty cycle after longer episodes of
sustained growth, when significant amounts of gas can be expelled in a
large-scale outflow from the innermost region of the host galaxy. 

Finally, we have studied the case of isolated binary major mergers of two disc
galaxies hosting black holes. Due to very efficient torquing gas angular
momentum is efficiently transported outwards and large amounts of gas are
funnelled towards the centre of the merger remnant. With respect to the
isolated disc galaxy model, the growth of the black hole using the vorticity
prescription is greatly increased as expected, but it still remains
significantly suppressed when compared to merger remnant grown with the
standard Bondi rate. Interestingly, while the final black hole mass in the
merger runs with and without vorticity suppression is similar, there is a
several Gyr delay in reaching this mass, once the gas angular momentum barrier
is taken into account. This could naturally explain scarce observational
evidence of quasar triggering in galaxies with perturbed morphologies which
previously had been accounted for by obscuration effects alone. Indeed, 
there is observational evidence of a significant delay between starburst activity and the peak
of AGN activity \citep[e.g.][]{Davies:07,Bennert:08,Yesuf:14,Matsuoka:15}.

We finally caution that our work is just a first stab in the direction of
understanding the interplay between the black hole growth and the gas angular
momentum from the point of view of galaxy formation simulations. As we have
demonstrated in this work, the black hole accretion suppression within
our model is highly sensitive on the gas properties on parsec scales. Thus
a more realistic treatment of the ISM physics and black hole feedback are needed
to make more progress on this front.  More generally, the amount of gas that
may eventually reach the black hole will be heavily modulated by the physics
occurring on 
even smaller spatial spaces, where effectiveness of gas fragmentation, star
formation (and stellar winds) as well as (magneto-) hydrodynamical viscous
transport processes through the accretion disc need to be considered. While
numerically self-consistently accounting for all these processes on such a
vast range of spatial scales will be a formidable challenge for some time to
come, 
simulations have now 
started to reach the regime where some of these scales can be bridged,
promising to shed light on the physics that powers supermassive black
hole growth. 

\section{Acknowledgements}
We thank Ewald Puchwein, Martin Haehnelt and Volker Springel for their useful comments on our manuscript. We thank Ruediger Pakmor for proving us with his initial condition setup for the
Keplerian disc simulation. MC is supported by the Science and
Technology Facilities Council (STFC). DS acknowledges support by the STFC and
the ERC Starting Grant 638707 ``Black holes and their host galaxies:
co-evolution across cosmic time''. This work was performed on the following: the COSMOS Shared Memory system at DAMTP,
University of Cambridge operated on behalf of the STFC DiRAC HPC
Facility - this equipment is funded by BIS National E-infrastructure capital
grant ST/J005673/1 and STFC grants ST/H008586/1, ST/K00333X/1; DiRAC Darwin
Supercomputer hosted by the University of Cambridge
High Performance Computing Service
(http://www.hpc.cam.ac.uk/), provided by Dell Inc.
using Strategic Research Infrastructure Funding from
the Higher Education Funding Council for England and
funding from the Science and Technology Facilities Council; DiRAC Complexity
system, operated by the University of Leicester IT Services. This equipment is
funded by BIS 
National E-Infrastructure capital grant ST/K000373/1 and
STFC DiRAC Operations grant ST/K0003259/1; COSMA Data Centric system at Durham
University, operated by the Institute for Computational Cosmology on behalf of
the STFC DiRAC HPC Facility. This equipment was funded by a BIS National
E-infrastructure capital grant ST/K00042X/1, STFC capital grant ST/K00087X/1,
DiRAC Operations grant ST/K003267/1 and Durham University. DiRAC is
part of the National E-Infrastructure.

\bibliographystyle{mn2e} 
\bibliography{references}

\appendix
\label{appendix}
\section{Setting the $f_\mathrm{sort}$ threshold}
\begin{figure}
\centering
\includegraphics[width=0.99\columnwidth]{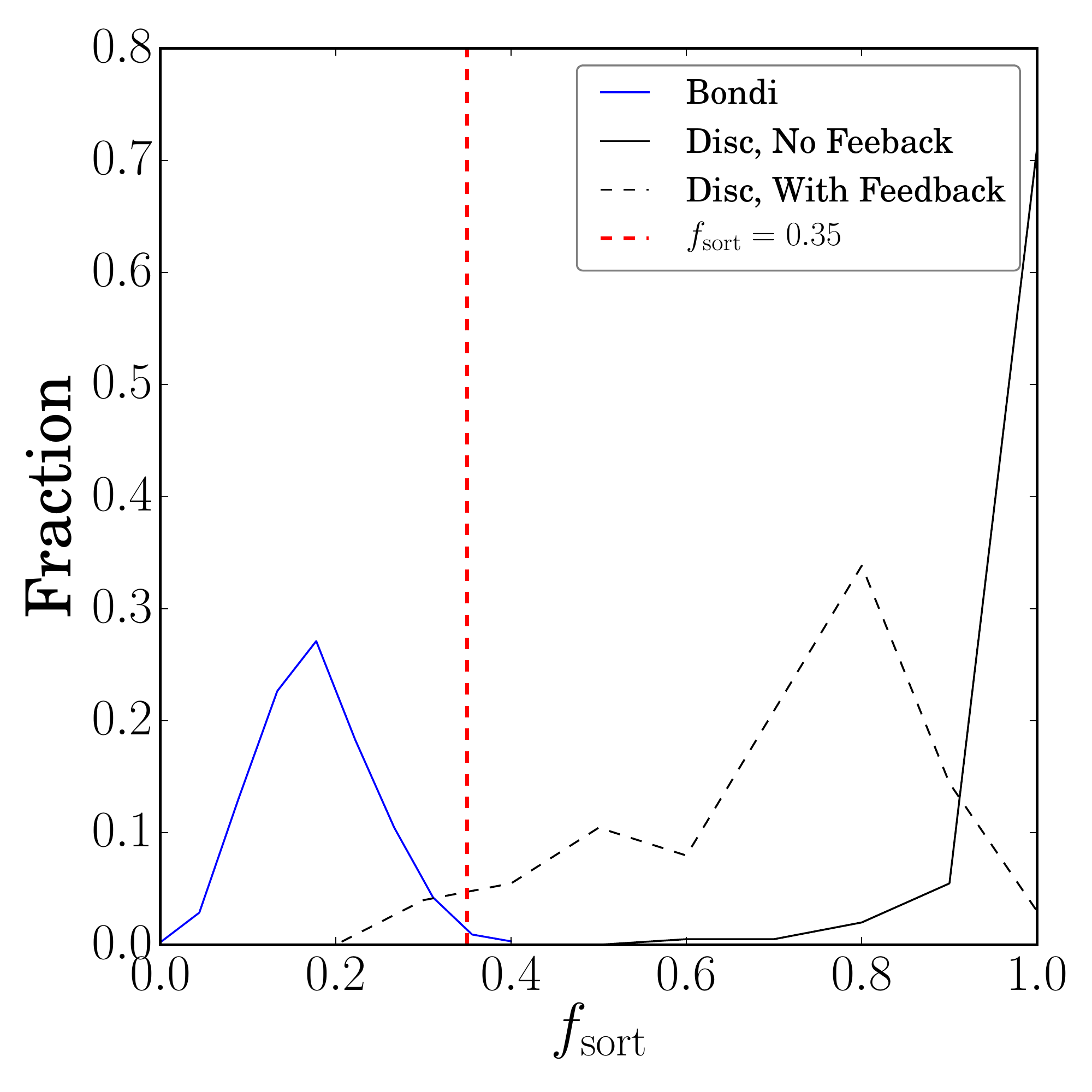}
\caption{The distribution of $f_\mathrm{sort}$ for simulations of Bondi flow, as well as simulations of isolated disc galaxies with and without feedback.}
\label{fsort} 
\end{figure}

As discussed in Section \ref{spurious_vorticity}, $f_\mathrm{sort}$ is the fraction of the gas mass that has a vorticity within $\uppi/4$ of the direction of the net vorticity. We set a minimum value that represents the threshold for using our vorticity prescription.

Figure~\ref{fsort} shows the distribution of the fraction of gas which is aligned with the net vorticity for simulations of Bondi flow, as well as for simulations of isolated disc galaxies with and without feedback. In the case of the Bondi simulation, the vorticity signal is noise driven and so the fraction aligned with the net vorticity follows that expected for a random distribution. In the case of the disc galaxy, the signal is coherent, especially when there are no perturbations to the velocity field caused by feedback. We set the threshold at which the vorticity suppression is active to be $f_\mathrm{sort} = 0.35$, to exclude the scenario present in the Bondi simulations.

\end{document}